\documentclass[12pt]{iopart}
\usepackage{amssymb}
\usepackage{epsf}
\usepackage{float}

\begin{document}
\jl{19}

\title{Dynamical correlations and collective excitations of Yukawa liquids}

\author{Z. Donk\'o$^{1,2}$\footnote[2]{To
whom correspondence should be addressed (donko@mail.kfki.hu)}, G. J.
Kalman$^2$, P. Hartmann$^{1,2}$}

\address{$^1$Research Institute for Solid State Physics and Optics,\\
Hungarian Academy of Sciences, POB 49, H-1525 Budapest, Hungary}

\address{$^2$Department of Physics, Boston College, 
Chestnut Hill, MA 02467, USA}

\date{\today}

\begin{abstract}

In dusty (complex) plasmas, containing mesoscopic charged grains,
the grain-grain interaction in many cases can be well described
through a Yukawa potential. In this Review we summarize the basics of
the computational and theoretical approaches capable of describing
many-particle Yukawa systems in the liquid and solid phases and
discuss the properties of the dynamical density and current
correlation spectra of three- and two-dimensional strongly coupled
Yukawa systems, generated by molecular dynamics simulations. We show
details of the $\omega(k)$ dispersion relations for the collective
excitations in these systems, as obtained theoretically following the
quasilocalized charge approximation, as well as from the fluctuation
spectra created by simulations. The theoretical and simulation results
are also compared with those obtained in complex plasma experiments.

\end{abstract}


\section{Introduction}

Strongly coupled plasmas -- in which the average potential energy per
particle dominates over the average kinetic energy -- appear in a wide
variety of physical systems: dusty plasmas, charged particles in
cryogenic traps, condensed matter systems such as molten salts and
liquid metals, electrons trapped on the free surface of liquid helium,
astrophysical systems, such as the ion liquids in white dwarf
interiors, neutron star crusts, supernova cores, and giant planetary
interiors, as well as in degenerate electron or hole liquids in
two-dimensional or layered semiconductor nanostructures
\cite{SCCS}. Many of these systems share some properties, which allows
modeling them by considering explicitly only a single type of charged
species and using a potential that accounts for the presence and
effects of other types of species. This latter may be thought of as
forming a charge-neutralizing background, which is either
non-polarizable or polarizable. In the first case the interaction of
the main plasma constituents can be expressed by the $\phi(r) \propto
1/r$ Coulomb potential, while in the case of polarizable background
the use of the $\phi(r) \propto \exp(-r/\lambda_{\rm D})/r$ Yukawa
potential is appropriate to account for screening effects
($\lambda_{\rm D}$ is the Debye length). Perhaps the most important
realizations of systems lending themselves to the approximation of the
interaction by the Yukawa potential are charged colloids
\cite{c1a,c1b,c2,c3} and dusty (complex) plasmas \cite{Konopka00} (for
comprehensive review on dusty plasmas see
e.g. \cite{FortovPR05,Ishihara07}).

In the case of 2D colloidal systems the microscopic particles move
in thin liquid films or between two closely separated glass plates.

In dusty plasmas both three-dimensional (3D) and two-dimensional (2D)
settings appear in nature and in laboratory environments. In
laboratory experiments 2D systems appear as a layer of dust particles
levitated in gaseous discharges. While most of the studies on this
latter system have been carried out in the crystalline state (for
early references to ``plasma crystals'' see
\cite{crystals1,crystals2,crystals3,Thoma04}), the liquid state is
receiving more current attention
\cite{transport1,transport2,Nunomura05,Piel06,Nosenko06,Chan2005}.
The important difference between colloidal and dusty plasma systems is
in the damping rate in particle dynamics and concomitantly in the wave
dynamics. In colloidal suspensions the background liquid exerts a
large friction on the moving charged particles, while in dusty plasmas
the background is gaseous and therefore the friction is lower and the
damping of the waves is weak. For this reason our focus in this Review
will be directed at dusty plasmas. The Review will cover studies of
strongly coupled plasmas mostly in the {\it liquid state}, where both
free motion and localization intervene.  The principal observation is
that from the point of view of collective behavior it is the
localization -- even though imperfect localization, or
\textit{quasilocalization} -- of particles that plays the principal
role. In contrast to the Vlasov plasma where the collective modes
arise from the fluid-like continuum behavior, in the strongly coupled
liquid they are more related to the normal modes of the interacting
quasilocalized particles. This, of course, suggests a link with the
harmonic phonon theory of crystal lattices.  At the same time, one has
to allow for the randomness of the distribution of the particles and
for the finite lifetime of the localization in the constantly changing
potential landscape. This latter process is expected to be primarily
responsible for the damping of the collective modes, in contrast both
to Vlasov plasmas, where Landau damping dominates and to weakly
correlated plasmas where collisional damping is the principal damping
mechanism. This physical picture suggests a microscopic
equation-of-motion model where the particles are trapped in local
potential fluctuations. The particles occupy randomly located sites
and undergo oscillations around them. At the same time, however, the
site positions also change and a continuous rearrangement of the
underlying quasi-equilibrium configuration takes place. Inherent in
this description is the assumption that the two time scales are well
separated and that for the description of the fast oscillating motion,
the time average (converted into ensemble average) of the drifting
quasi-equilibrium configuration is sufficient. Here the distinction
between the "direct" and "indirect" thermal effects should be
emphasized: the former are responsible for the actual motion and
migration of the particles, the latter refer to the accessibility of
the possible configurations of the random sites and to the temperature
dependence of the probability of a particular configuration.

The interaction potential energy of particles in \textit{Yukawa
liquids} is given by (e.g. \cite{SH_PI}):
\begin{equation}
\phi(r)=\frac{Q^2}{4 \pi \varepsilon_0}{{\exp( - r / \lambda_{\rm D})} \over
{r}}, \label{eq:yukawa}
\end{equation}
where $Q$ is the charge of the particles, $\varepsilon_0$ is the
permittivity of free space, and the Debye length
$\lambda_{\rm D}$ accounts for the screening of the interaction by
other plasma species. The main (dimensionless) parameters, which
fully characterize the systems are: (i) the {\it coupling
parameter} (defined in the same way as for Coulomb systems):
\begin{equation}\label{eq:gamma}
    \Gamma = \frac{Q^2}{4 \pi \varepsilon_0} \frac{1}{a k_{\rm B} T},
\end{equation}
where $a$ is the Wigner-Seitz (WS) radius and $T$ is the
temperature, and (ii) the {\it screening parameter}
\begin{equation}\label{eq:kappa}
    \kappa = \frac{a}{\lambda_{\rm D}}.
\end{equation}
$\Gamma$ is the customary measure of the ratio of the average
potential energy to the average kinetic energy per particle; the
strong coupling regime, relevant here, corresponds to $\Gamma \gg$
1. In the $\kappa \rightarrow 0$ limit the interaction reduces to
the Coulomb type, while at $\kappa \rightarrow \infty$ it
approximates the properties of a hard sphere interaction.

At $\kappa$ = 0, in 3D, the liquid domain is limited to coupling
parameter values $\Gamma \lesssim 175$
\cite{Slattery80,Slattery82,Farouki93}, where the plasma is known to
crystallize into a bcc lattice \cite{brush1966,stringfellow1990}. In 2D,
crystallization into hexagonal lattice occurs at a lower value of
coupling, at $\Gamma \approxeq 137$, as found by computer simulations
\cite{muto99,gann1979} and proven by experiments \cite{Grimes1} as
well. At $\kappa >$ 0, 3D systems may crystallize either in a bcc or in
a fcc lattice, depending on $\kappa$, as found by Hamaguchi {\it et al.}
\cite{hamaguchi1997} in their calculation of the phase diagram of
Yukawa systems. In 2D the crystallized form of the systems is always
hexagonal.

The possibility of characterizing a Yukawa liquid with an effective
coupling coefficient $\Gamma^\star$ instead of a ($\Gamma,\kappa$)
parameter pair has recently been addressed. In 3D $\Gamma^\star$ was
derived by Vaulina, Fortov and co-workers
\cite{vaulina2002a,vaulina2002b,fortov2003} on the basis of the
frequency of dust lattice waves. Subsequently, for 2D Yukawa liquids
the $\Gamma^\star = f(\Gamma,\kappa)$ relationship was established by
prescribing a constant amplitude for the first peak of the pair
correlation function for fixed values of $\Gamma^\star$
\cite{H2005}. The solid-liquid melting line ($\Gamma_{\rm melting}$
vs. $\kappa$) in both 3D and 2D system was found in these studies to
follow closely constant $\Gamma^\star$ values.

Additional characteristic parameters of the systems investigated here
are the WS radius $a$ and the plasma frequency $\omega_0$, which are
given for 3D and 2D systems by:
\begin{eqnarray}
a_{\rm 3D} = (4 n_{\rm 3D} \pi /3)^{-1/3} \label{eq:aws3d} \\
a_{\rm 2D} = (n_{\rm 2D} \pi)^{-1/2}, \label{eq:aws2d}
\end{eqnarray}
where $n_{\rm 3D}$ and $n_{\rm 2D}$ are the 3D (number) density and the 2D
areal (number) density of particles, and
\begin{equation} \omega_{\rm 0, 3D} =
\sqrt{\frac{n_{\rm 3D} Q^2}{\varepsilon_0 m}} \label{eq:omegap3d}
\end{equation}
and
\begin{equation}
\omega_{\rm 0, 2D} = \sqrt{\frac{n_{\rm 2D} Q^2}{2 \varepsilon_0 m a_{\rm
2D}}}. \label{eq:omegap2d}
\end{equation}
Note that in 2D the nominal plasma frequency $\omega_0$ may also have
different definitions, and some of the authors use the lattice
constant instead of the WS radius as a length scale.

Other important frequencies characterizing strongly coupled plasmas
are the Einstein frequencies, which are the normal modes of
oscillation of a test charge in the presence of a given (static)
distribution of charges. Einstein frequencies are well known for
lattice structures, however, there has been relatively little work
done on disordered and liquid phase systems \cite{Bakshi}. Such
systems are being studied through the combination of theoretical and
simulation approaches \cite{DKG_caging,DHK_aps}.

As to the possibilities of theoretical description, many-body systems
can be treated theoretically in a straightforward way in the extreme
limits of both weak interaction and very strong interaction. In the
first case, one is faced with a gaseous system, or a Vlasov plasma,
where correlation effects can be treated perturbatively ($\Gamma \ll
1$). In the case of very strong interaction, the system crystallizes,
the particles are completely localized and phonons are the principal
excitations. In the intermediate regime -- in the strongly coupled
liquid phase -- the localization of the particles in the local minima
of the potential surface still prevails, however, due to the diffusion
of the particles the time of localization is finite
\cite{DKG_caging}. The localization of the particles (which may
typically cover a period of several plasma oscillation cycles) serves
as the basis of the \emph{Quasi-Localized Charge Approximation} (QLCA)
method \cite{qlca1,qlca2}. Besides the theoretical approaches computer
simulations have proven to be invaluable tools for investigations of
strongly coupled liquids of charged particles. Monte Carlo (MC) and
molecular dynamics (MD) methods have widely been applied in studies of
the equilibrium and transport properties, as well as of dynamical
effects and collective excitations. The main difference between the
two techniques is that in an MC simulation the particle configuration
with the lowest energy is searched for, whereas MD simulations provide
information about the time-dependent phase space coordinates of the
particles, this way allowing studies of dynamical properties.

This paper intends to review the dynamical properties and collective
behavior of strongly coupled Yukawa systems in the liquid and solid
phases, in two and three dimensions. First we describe the numerical
as well as theoretical methods used, in Sections \ref{sec:MD} and
\ref{sec:theory}, respectively.  The analysis of the collective mode
behavior of 3D liquids is presented in Section \ref{sec:threeD}. In
the 2D case we investigate both an ideally narrow particle layer, and
a layer having a finite width, where particles are confined by an
external parabolic potential. The analysis of these systems is
described in Sections \ref{sec:twoD} and \ref{sec:qtwoD},
respectively.  Section \ref{sec:experiment} gives a brief summary of
the experimental studies relevant to the theoretical and simulation
results reviewed. Section \ref{sec:sum} gives a summary of the paper
as well as a short outlook on the topics, which may have been
additional subjects of this Review.

\section{Molecular dynamics simulations}\label{sec:MD}

Molecular dynamics simulations follow the motion of particles by
integrating their equations of motion while accounting for the
pairwise interaction of the particles, as well as for the forces originating from any external field(s), see e.g. \cite{frenkel}.
In the plasma / gas background environment friction forces and
randomly fluctuating forces also act on the particles in
addition to the forces arising from the interaction of the
electrically charged particles. The general form of the
equation of motion (of a ``testÕÕ particle $i$) is (see e.g. \cite{B2006}):
\begin{equation}\label{eq:eom}
    m \ddot{\bf r}_i = \sum_{i \ne j} {\bf F}_{i,j}(t) + {\bf F}_{\rm ext}(t)
    - m \eta {\bf v}_i(t) + {\bf R},
\end{equation}
where ${\bf F}_{i,j}$ is the force originating from the interaction
with particle $j$, ${\bf F}_{\rm ext}$ is the force originating from
any external field, $\eta$ is the friction coefficient, and ${\bf R}$
represents a Brownian randomly fluctuating force (Langevin force).
The results presented here correspond to ``idealized'' Yukawa liquids
for which $\eta=0$ and ${\bf R}=0$ are assumed. Also, in most of our
studies we investigate infinite (unconfined) systems, for which ${\bf
F}_{\rm ext}=0$, as well, although as an example a
quasi-two-dimensional liquid -- confined by an external parabolic
field -- is also studied. In the case of unconfined systems periodic
boundary conditions (PBC) are imposed in the simulations. In the case
of confinement along one of the coordinates, PBC-s are used in the
unconfined directions. It is noted that for charged colloids Brownian
molecular dynamics simulation \cite{c1b,HL92} is widely used, which
represents the extreme limit of large friction and large ${\bf R}$. In
this case the inertial term ($m \ddot{\bf r}_i$) in eq. (\ref{eq:eom}) 
is neglected.

The calculation of the force acting on a particle of the system, ${\bf
F}_i$, is relatively simple in the case of short-range potentials
(like the Lenard-Jones potential or the $1/r^{12}$ potential).  In
this case MD methods make use of the truncation of the interaction
potential thereby limiting the need for the summation of pairwise
interactions around a test particle to a region of finite size. In the
case of long-range interactions (e.g. Coulomb or low-$\kappa$ Yukawa
potentials), which are also of interest here, however, such truncation
of the potential is not allowed, and thus special techniques, like
Ewald summation \cite{Ewald}, have to be used in MD
simulations. Besides the Ewald summation technique there exist few
additional methods, like the fast multipole method and the
particle-particle particle-mesh method (PPPM, or P3M), which can be
used to handle long-range interaction potentials, see e.g.
\cite{Sagui1999}. The results presented here for Coulomb systems have
been obtained from simulations using this latter method
\cite{Hooker,Gargallo,David,EHL80,HE81}.  In the PPPM scheme the
interparticle force is partitioned into (i) a force component ${\bf
F}_{\rm PM}$ that can be calculated on a mesh (the ``mesh force'') and
(ii) a short-range (``correction'') force ${\bf F}_{\rm PP}$, which is
to be applied to closely separated pairs of particles only. In the
mesh part of the calculation charged clouds are used instead of
point-like particles and their interaction is calculated on a
computational mesh, taking also into account periodic images (for more
details see \cite{EHL80,HE81}). This way the PPPM method makes it
possible to take into account periodic images of the system (in the PM
part), without truncating the long range Coulomb or low-$\kappa$
Yukawa potentials. For screening values $\kappa \gtrsim 1$ the PP part
alone provides sufficient accuracy. In these cases the mesh part of
the calculation is not used, the interaction forces are summed for
particles situated within a ($\kappa$-dependent) cutoff radius around
the test particle. Identification of these ``neighboring'' particles
is aided by the ``chaining mesh technique''.

In the simulations presented here usually a spatially random particle
configuration is set up at the initialization, with particle
velocities sampled from a Maxwellian distribution of temperature
$T_0$, which corresponds to the desired value of the coupling
parameter $\Gamma$ [see eq.~(\ref{eq:gamma})].  The equations of
motion of the particles are integrated using the leapfrog scheme or
the velocity-Verlet scheme. The desired system temperature is reached
by rescaling the particle momenta during an initialization phase of
the simulation. In equilibrium MD simulations measurements on the
system are taken following this phase, in the state of thermodynamic
equilibrium. During this phase thermostation is usually no longer
applied. If thermostation is necessary, rescaling of particle
velocities is to be avoided, algorithms like the Nos\'e-Hoover
thermostat can be applied (see e.g. \cite{frenkel,nose,hoover}).

In our studies measurements on the system are taken
at constant volume ($V$), particle number ($N$) and total energy
($E$). The MD simulations directly provide the pair correlation
function (PCF) of the system, which is the basis for the calculation
of thermodynamic quantities (not detailed here, see e.g.
\cite{H2005}), and is also required as input to the QLCA equations
for the calculation of the dispersion relations and other quantities
(see later).

In the MD simulation information about the (thermally excited)
collective modes and their dispersion is obtained from the Fourier
analysis of the correlation spectra of the density fluctuations
\begin{equation}\label{eq:rho}
\rho(k,t)= k \sum_j \exp \bigl[ i k x_j(t) \bigr]
\end{equation}
yielding the dynamical structure function as \cite{HMP75}:
\begin{equation}\label{eq:sp1}
S(k,\omega) = \frac{1}{2 \pi N} \lim_{\Delta T \rightarrow \infty}
\frac{1}{\Delta T} | \rho(k,\omega) |^2,
\end{equation}
where $\Delta T$ is the length of data recording period and
$\rho(k,\omega) = {\cal{F}} \bigl[ \rho(k,t) \bigr]$ is the
Fourier transform of (\ref{eq:rho}).

Similarly, the spectra of the longitudinal and transverse current
fluctuations, $L(k,\omega)$ and $T(k,\omega)$, respectively, can
be obtained from Fourier analysis of the microscopic quantities
\begin{eqnarray}
\lambda(k,t)= k \sum_j v_{j x}(t) \exp \bigl[ i k x_j(t) \bigr], \nonumber \\
\tau(k,t)= k \sum_j v_{j y}(t) \exp \bigl[ i k x_j(t) \bigr],
\label{eq:dyn}
\end{eqnarray}
where $x_j$ and $v_j$ are the position and velocity of the $j$-th
particle. Here we assume that ${\bf k}$ is directed along the $x$
axis (the system is isotropic) and accordingly omit the vector
notation of the wave number. The way described above for the
derivation of the spectra provides information for a series of wave
numbers, which are multiples of $k_{\rm min} = 2 \pi / H$, where $H$
is the edge length of the simulation box. The collective modes are
identified as peaks in the fluctuation spectra. The widths of the
peaks provide additional information about the lifetimes of the
excitations: narrow peaks correspond to longer lifetimes, while
broad features are signals for short lived excitations.

\section{Theoretical approaches}\label{sec:theory}

The Molecular Dynamics calculations compute the dynamical 
density--density and current--current correlations (dynamical structure
functions), from whose behavior the dispersion relations for the
collective modes can be inferred. Following the same route in a
theoretical analysis would be an extremely ambitious
undertaking. Calculating the dynamical structure functions is not an
easy task and not much progress has been achieved so far along this line.
The single-particle and collective microscopic dynamics of a classical
3D Yukawa fluid was first analyzed by Barrat {\it et al.} 
\cite{Barrat1998},
on the basis of memory function and mode-coupling theories. They have
found that the longitudinal current fluctuations and the velocity
autocorrelation function cross over continuously from the behavior
characteristic of classical fluids with short-range interactions to
the dynamics of a one-component plasma as the screening parameter
$\kappa$ of the Yukawa potential is reduced.

2D Yukawa systems in the liquid phase were considered by L\"{o}wen
\cite{HL92} and Murillo and Gericke \cite{MG03}. In this latter work
radial distribution functions have been computed with the hypernetted
chain equations and were compared with those obtained from molecular
dynamics simulations. The dynamical structure function obtained from
the RPA approach extended by local field corrections was shown to be
inadequate to reproduce the features of the structure function
obtained from molecular dynamics. Ref. \cite{HL92} focused mostly on
the static properties and Brownian dynamics of the system, while also
considering some features of the dynamical fluctuations. Applying the
viscoelastic approximation Murillo \cite{Murillo3} analyzed some
aspects of the transverse current fluctuations.

Fortunately, for the determination of the collective mode spectrum a
much more direct approach, via the analysis of the dielectric response
(tensor) function, is available. Thus the primary goals of the
analytical methods discussed below are the determination of the
dielectric function and the derivation of the ensuing dispersion
relation for the collective modes.

The dielectric tensor in the spatially homogeneous liquid phase is
diagonal in the coordinate system, where ${\bf k}$ is along one of the
coordinate axes. Accordingly, the collective modes can be classified
by their polarization into {\it longitudinal} and {\it transverse}
modes. In the crystalline solid phase the rotational symmetry is
broken, the structure of the dielectric tensor is more intricate and
the longitudinal and transverse polarizations do not, in general,
represent eigenpolarizations anymore. In this Review we are concerned
with the collective mode structure of the liquid phase, but the
understanding of the behavior of the collective modes in the solid
phase has a bearing, as we will discuss, on the formation of the
collective modes in the strongly couple liquid phase as well.

\subsection{Fluctuation--Disspation Theorem}

The link between the $S({\bf k},\omega)$, $L({\bf k},\omega)$, $T({\bf
  k},\omega)$ spectra measured in the simulations and the dielectric
function is provided by the Fluctuation--Dissipation Theorem
\begin{eqnarray}
S({\bf k},\omega)&=&\frac{k^2}{\omega^2}L({\bf k},\omega)
=\frac{1}{\pi\beta n \omega}{\rm Im} \chi_{\rm L}({\bf k},\omega)
=-\frac{1}{\pi\beta n \omega} \frac{{\rm Im} \bar{\chi}_{\rm L}({\bf
    k},\omega)}{\left| \varepsilon_{\rm L}({\bf k},\omega) \right|^2},
\\ T({\bf k},\omega)&=&\frac{\omega^2}{k^2}\frac{1}{\pi\beta n
  \omega}{\rm Im} \chi_{\rm T}({\bf k},\omega), \nonumber
\end{eqnarray}
where $\beta = 1 / kT$, $\chi_{\mu\nu}( {\bf k},\omega)$ is the
susceptibility tensor, and $\bar{\chi}_{\mu\nu}( {\bf k},\omega)$ is the
proper (or total) susceptibility tensor.

At the $\Omega$ value where the dispersion relation is satisfied,
$\chi_{\rm L,T}^{-1}({\bf k},\Omega)$ vanishes. This, in general,
happens only at a complex frequency, the imaginary part of which being
characteristic of the damping of the mode. Since the dynamical
structure functions are plotted and analyzed for real frequencies
only, $\chi_{\rm L,T}^{-1}({\bf k},\Omega)$ reaches only a minimum at
some value of the real $\omega$, which can be expected to be in the
vicinity of the actual complex $\Omega$: this is the frequency value
that can be identified at which the peak of the fluctuation spectrum
occurs.

\subsection{Dielectric Response Function}

The tensorial dielectric response function $\varepsilon_{\mu\nu}( {\bf
  k},\omega)$ can be expressed either in terms of the susceptibility
tensor $\chi_{\mu\nu}( {\bf k},\omega)$ or the proper (or total)
susceptibility tensor $\bar{\chi}_{\mu\nu}( {\bf k},\omega)$ and the
Fourier transform $\varphi(k)$ of the interaction potential
(\ref{eq:yukawa}). This latter depends on the dimensionality of the
system. In 3D
\begin{equation}
  \varphi(k)=\frac{1}{\varepsilon_0}\frac{Q^2}{k^2+\kappa^2}
\end{equation}
and in 2D
\begin{equation}
  \varphi(k)=\frac{1}{2\varepsilon_0}\frac{Q^2}{\left(k^2+\kappa^2\right)^{1/2}}.
\end{equation}
Then
\begin{equation}
  \varepsilon_{\mu\nu}( {\bf k},\omega) = \delta_{\mu\nu} -
  \varphi(k)\chi_{\mu\nu}( {\bf k},\omega).
\end{equation}
In the coordinate system where ${\bf k}$ is along the $z$ axis in 3D and
along the $y$ axis in 2D, the isotropic liquid $\varepsilon_{\mu\nu}(
{\bf k},\omega)$ has the structure
\begin{eqnarray}
\varepsilon_{\mu\nu}( {\bf k},\omega) = \left[ \begin{array}{ccc}
\varepsilon_{\rm T}( {\bf k},\omega) & 0 & 0 \\ 0 & \varepsilon_{\rm
T}( {\bf k},\omega) & 0 \\ 0 & 0 & \varepsilon_{\rm L}( {\bf
k},\omega) \end{array} \right] &~~~{\rm 3D}\\ \varepsilon_{\mu\nu}(
{\bf k},\omega) = \left[ \begin{array}{cc} \varepsilon_{\rm T}( {\bf
k},\omega) & 0 \\ 0 & \varepsilon_{\rm L}( {\bf k},\omega) \end{array}
\right] &~~~{\rm 2D} \nonumber
\end{eqnarray}
and the dispersion relations for the collective modes are given by
\begin{eqnarray}\label{eq:dieldisp}
\varepsilon_{\rm L}( {\bf k},\omega)=0, &~~~~{\rm (a)} \\
\varepsilon^{-1}_{\rm T}( {\bf k},\omega)=0. &~~~~{\rm (b)} \nonumber
\end{eqnarray}

The longitudinal dielectric function has the immediate physical
significance that it relates the externally imposed electric field to
the total (external+polarization) field by $E_{\rm total}({\bf
  k},\omega) = E_{\rm external}({\bf k},\omega) / \varepsilon_{\rm L}(
  {\bf k},\omega)$. In contrast, the transverse dielectric function
has well-defined physical meaning only in terms of the full
electrodynamics of the system \cite{GKS_JSP}. Here, there is a certain
degree of arbitrariness in the definition of $\varepsilon_{\rm T}( {\bf
  k},\omega)$. A useful alternative formulation of the dispersion
relations is in terms of the external susceptibility
\begin{eqnarray}
\chi_{\rm L}( {\bf k},\omega) &=& \frac{\bar{\chi}_{\rm L}( {\bf
    k},\omega)}{\varepsilon_{\rm L}( {\bf k},\omega)}, \\
\chi_{\rm T}( {\bf k},\omega) &=& \bar{\chi}_{\rm T}( {\bf
    k},\omega). \nonumber
\end{eqnarray}
 This allows expressing the condition for the collective
excitation in the universal form
\begin{equation}
\chi_{\rm L,T}^{-1}( {\bf k},\omega)=0.
\end{equation}

$\bar{\chi}_{\mu\nu}( {\bf k},\omega)$ embodies all the dynamical
properties of the system, which stem partly from interparticle
correlations, partly from the random motion of the particles. Over the
past half century an immense effort has gone into the calculation of
this quantity for Coulomb systems, both classical and quantum.  Most
of the work focused on weakly coupled ($\Gamma \ll 1$) or moderately
coupled ($1<\Gamma<10$) systems. Interest in strongly coupled Coulomb
and Yukawa systems is more recent \cite{BausHansen, TosiMarch}. In the
strongly coupled domain the dynamics is dominated by
correlations. Here and in the sequel we will mostly ignore the effect
of thermal motions on $\chi_{\mu\nu}( {\bf k},\omega)$; some comments
on how to abandon this simplification will be made later in this
Section.

While our focus in this Review is on the strongly coupled liquid
phase, it will be instructive and of interest to begin with an
orientation based on the weakly coupled Random Phase
Approximation (RPA) theory. The RPA or Vlasov
description is based on the assumption that the mean field dominates
the particle-particle interaction and correlations can be
ignored. This is tantamount to taking $\bar{\chi}_{\mu\nu}( {\bf
k},\omega)$ as that of the non-interacting gas (although perhaps not
quite obviously: for a discussion see e.g. \cite{PinesNozieres}):
$\bar{\chi}_{\mu\nu}( {\bf k},\omega)=\chi_0( {\bf
k},\omega)\delta_{\mu\nu}$, which, with the neglect of thermal motion is
\begin{equation}
  \chi_0( {\bf k},\omega)=\frac{n}{m}\frac{k^2}{\omega^2}.
\end{equation}
This leads to the simple expressions for the elements of the
dielectric tensor
\begin{eqnarray}
\varepsilon_{\rm L}( {\bf k},\omega)=\varepsilon_{\rm T}( {\bf
  k},\omega)=1-\frac{\bar{k}^2}{\bar{k}^2+\kappa^2}
\frac{\omega^2_{\rm 0,3D}}{\omega^2},&~~~~{\rm 3D} \\ \varepsilon_{\rm L}( {\bf
  k},\omega)=\varepsilon_{\rm T}( {\bf
  k},\omega)=1-\frac{\bar{k}^2}{\left( \bar{k}^2+\kappa^2\right)^{1/2}}
\frac{\omega^2_{\rm 0,2D}}{\omega^2},&~~~~{\rm 2D} \nonumber
\end{eqnarray}
where $\omega_{\rm 0,3D}$ and $\omega_{\rm 0,2D}$ are the respective 3D plasma
frequency and the 2D nominal plasma frequency defined in
Eqs. (\ref{eq:omegap3d}) and (\ref{eq:omegap2d}); $\bar{k}=ka$.

The dispersion relations (and their small-$k$ approximations) for the
3D and 2D longitudinal modes follow immediately from
(\ref{eq:dieldisp}):
\begin{eqnarray}\label{eq:Omega0}
\Omega^2_{\rm 0,3D}({\bf k})=
\omega^2_{\rm 0,3D}\frac{\bar{k}^2}{\bar{k}^2+\kappa^2}
\approx\frac{\omega^2_{\rm 0,3D}}{\kappa^2}\bar{k}^2,&~~~~{\rm 3D}
  \\ \Omega^2_{\rm 0,2D}({\bf k})=
  \omega^2_{\rm 0,2D}\frac{\bar{k}^2}{\left(\bar{k}^2+\kappa^2\right)^{1/2}}
  \approx\frac{\omega^2_{\rm 0,2D}}{\kappa}\bar{k}^2.&~~~~{\rm 2D}
    \nonumber
\end{eqnarray}
For $k \rightarrow 0$ the longitudinal mode is {\it acoustic},
i.e. $\omega_{\rm L}(k \rightarrow 0)=sk$, with the 3D and 2D acoustic
  velocities $s_{\rm 3D}$ and $s_{\rm 2D}$:
\begin{eqnarray}
s_{\rm 3D}=\frac{\omega_{\rm 0,3D}}{\kappa}, \\
s_{\rm 2D}=\frac{\omega_{\rm 0,2D}}{\sqrt{\kappa}}. \nonumber
\end{eqnarray}
Note that if we compare 3D and 2D systems with the same average
interparticle distance, the acoustic speed in 2D is different by a factor
$\sqrt{\frac{2}{3}\kappa}$ than in 3D. The acoustic behavior in 2D is of
course at complete variance with the corresponding $k \rightarrow 0$
of an unscreened Coulomb plasma, i. e. the limit $\kappa=0$,
where $\omega \propto \sqrt{k}$.

It is clear that there is no mode satisfying the transverse dispersion
relation (\ref{eq:dieldisp}b): the mean field RPA model, which is
devoid of correlations, cannot support a transverse shear wave, since
shear is a fundamentally correlational phenomenon.

\subsection{Quasi Localized Charge Approximation}\label{sec:qlca}

While the RPA provides a description of the weakly coupled gas, the
strongly coupled liquid state of a Coulombic or Yukawa system requires
a different approach. There have been various attempts over the years
to calculate dispersion relations and related quantities for such
systems. Noteworthy approaches include the high frequency sum rule
method \cite{HMP75}, the application of the STLS (Singwi, Tosi, Land
and Sjolander) technique originally developed for the electron gas in
metals \cite{Murillo1,Murillo2}, the memory function approach
\cite{Baus1,Baus2} and the viscoelastic model
\cite{Murillo3,Sen1,Ichimaru,Kaw2}.

In the long run, from a practical perspective most of these methods
have turned out to be problematic. The problems that occur vary: they
range from weak theoretical foundation through being more appropriate
for static than dynamical processes, to resulting in an unwieldy
formalism. On the other hand, a method originally proposed by Kalman
and Golden in \cite{KG90} that has become known as the Quasilocalized
Charge Approximation (QLCA) (for a review see \cite{qlca1, qlca2}) has
led to quite a successful history of accomplishments. The measure of
success in this context is (a) the ability to calculate from available
{\it static} data {\it dynamical} quantities that lend themselves to
comparison with numerical or laboratory experiments; (b) solid
agreement with the outcomes of MD simulations; and (c) a good accord
with the newly available laboratory experiments (still in a rather
limited number) on complex plasma wave propagation. The following is a
concise description of the QLCA method; for more details the reader is
referred to \cite{qlca1}.

The conceptual basis for the QLCA has been a model that implies the
following assumptions about the behavior of a strongly coupled Coulomb
or Yukawa liquid: (i) in the potential landscape within the many-body
system deep potential minima form that are capable of trapping
(caging) charged particles; (ii) a caged charge oscillates with a
frequency that is determined both by the local potential and the
interaction with the other (caged) particles in their instantaneously
frozen positions; (iii) the potential landscape changes slowly to
allow the charges to execute a fair number of oscillations; (iv) the
escape from the cages of the particles is caused by the gradual
disintegration of the caging environment; the time scale of this
process is governed by the coupling strength $\Gamma$; (v) the (time
and velocity dependent) correlation between a selected pair of
particles is well approximated by the (time and velocity independent)
equilibrium pair correlation; (vi) the frequency spectrum calculated
from the averaged (correlated) distribution of particles represents,
in a good approximation, the average over the distribution of
frequencies originating from the actual ensemble.

Hypotheses (i)--(iv) have undergone careful testing by a series of MD
simulation experiments both for Coulomb and Yukawa systems, and both
for 2D and 3D configurations \cite{DKG_caging,DHK_aps}, which will be
discussed in Section \ref{sec:simulation}. The validity of hypothesis
(v) has recently been called into question in relation to
multicomponent systems. The short time evolution of the pair
correlation function in the vicinity of a particle moving with respect
to its environment can certainly be velocity dependent and
anisotropic: it is now believed that it is this behavior that is
responsible for some discrepancies between MD simulation results and
QLCA predictions occurring in binary Coulomb and Yukawa systems. It is
not believed, however, that this behavior would be problematic in a
single component system. As to item (vi), the question of the
dynamical frequency distribution in a liquid has received very little
attention, either theoretically or experimentally (the record is
better in relation to disordered crystals, where the problem has been
posed and approximation schemes have been proposed, although in a
language where the central role of the dispersion relation is
obscured -- see, e.g. \cite{Elliot}). The extension of the QLCA in this
direction, while less than pressing, would be desirable.

The central quantity in the QLCA is the dynamical matrix, either in
three dimensions (${\cal{D}}=3$) or in two dimensions (${\cal{D}}=2$):
\begin{equation}
  D_{\mu \nu}({\bf k})= - \frac{n}{m} \int {\rm d}^{\cal{D}} r M_{\mu \nu}(r) 
  [{\rm e}^{i {\bf k} \cdot {\bf r}} -1] h(r), \label{eq:qlcadyna}
\end{equation}
which is formally similar to the eponymous quantity in the harmonic
theory of lattice phonons and is derived from the equation of motion
of properly constructed collective coordinates. $ M_{\mu
  \nu}(r)=\partial_{\mu} \partial_{\nu} \phi(r)$ is the dipole-dipole
interaction potential associated with $\phi(r)$. $D_{\mu
  \nu}({\bf k})$ is a functional of the equilibrium pair correlation
function (PCF) $h(r)$, or of its Fourier transform $h({\bf k})$.

The longitudinal and a transverse elements of the dielectric tensor
are now expressed in terms of corresponding elements of $D_{\mu
  \nu}({\bf k})$ :
\begin{equation}
\varepsilon_{\rm L/T}({\bf k},\omega)=
1-\frac{\Omega_{\rm 0}^2({\bf k})}{\omega^2-D_{\rm
    L/T}({\bf k})}.
\label{eq:qlcaeq2}
\end{equation}
Thus the $D_{\rm{L}}({\bf k})$ and $D_{\rm{T}}({\bf k})$ local field
functions are the respective projections of $D_{\mu \nu}({\bf k})$
\cite{qlca1}. $\Omega_0({\bf k})$ is the 3D or 2D longitudinal mode
frequency, found in (\ref{eq:Omega0}). One should keep
in mind that in spite of the universality of the expression
(\ref{eq:qlcadyna}) the explicit forms of the 3D and 2D
$\varepsilon_{\rm L}({\bf k},\omega)$-s are quite different.

We note that the input required in the calculations is the static
pair correlation function (PCF). In earlier works PCF-s generated
by the HNC (hypernetted chain \cite{L78}) technique have been used
as input data of the QLCA formulae to calculate the dispersion
relations. With the advent of computer simulation techniques it
turned out to be both more expedient and more accurate to import
simulation generated PCF-s in the theoretical calculations. The
results of the theoretical calculations presented in this paper use
PCF-s derived from molecular dynamics (MD) computations.

\begin{figure}[h!]
\begin{center}
\epsfxsize=13cm
\epsfbox{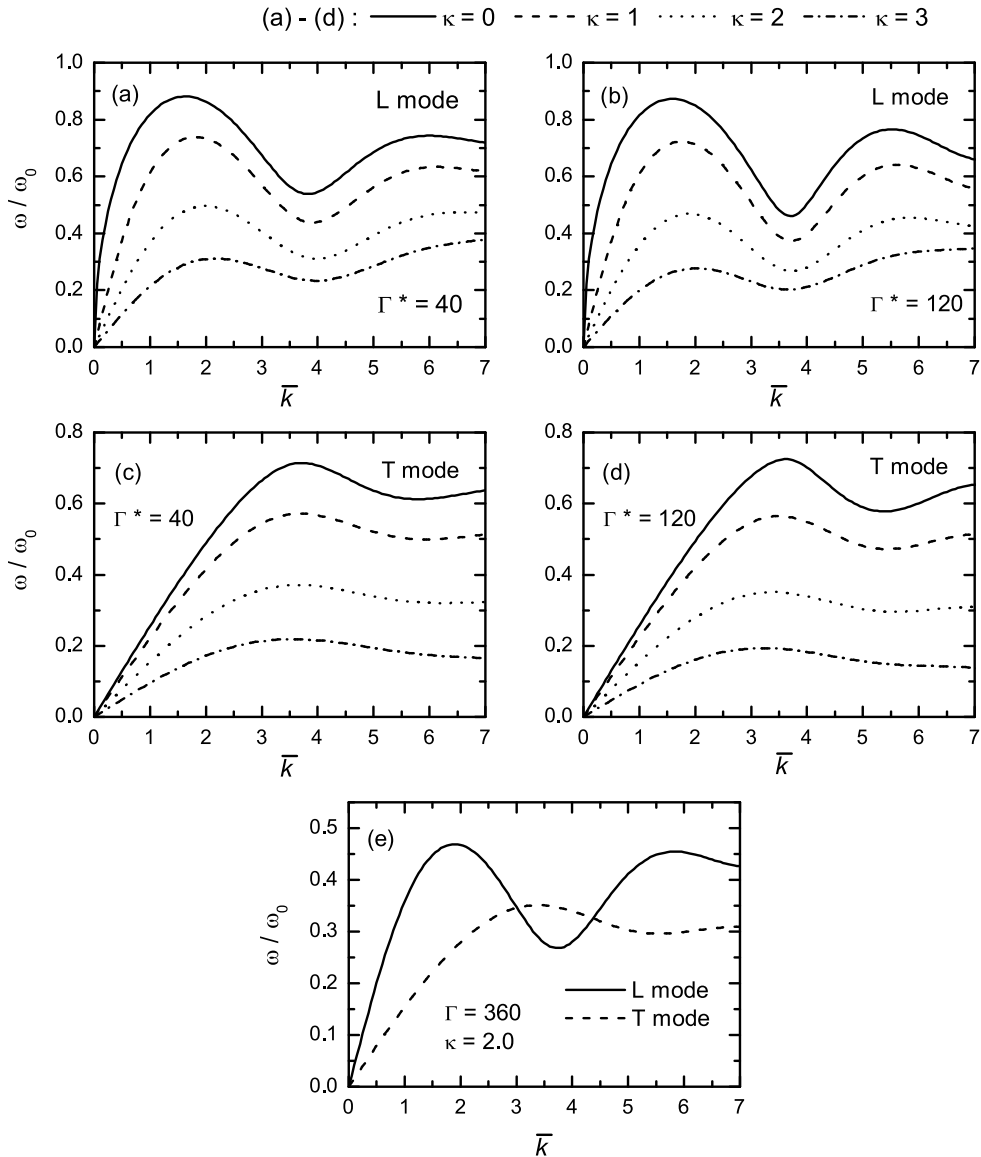}
\end{center}
\caption{2D Yukawa and Coulomb liquids: QLCA longitudinal and transverse
dispersions for specified values of the effective coupling
$\Gamma^\star$.}
\label{fig:tfig01}
\end{figure}

\begin{figure}[h!]
\begin{center}
\epsfxsize=13cm
\epsfbox{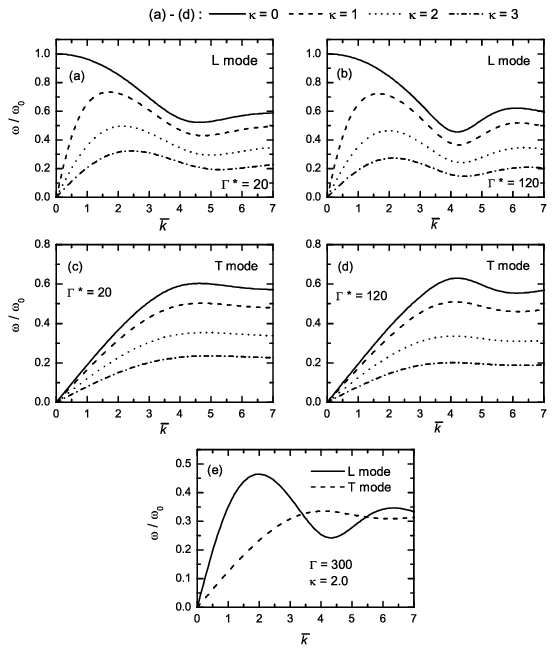}
\end{center}
\caption{3D Yukawa and Coulomb liquids: QLCA longitudinal and transverse
dispersions for specified values of the effective coupling
$\Gamma^\star$.}
\label{fig:tfig02}
\end{figure}

\begin{figure}[h!]
\begin{center}
\epsfxsize=13cm
\epsfbox{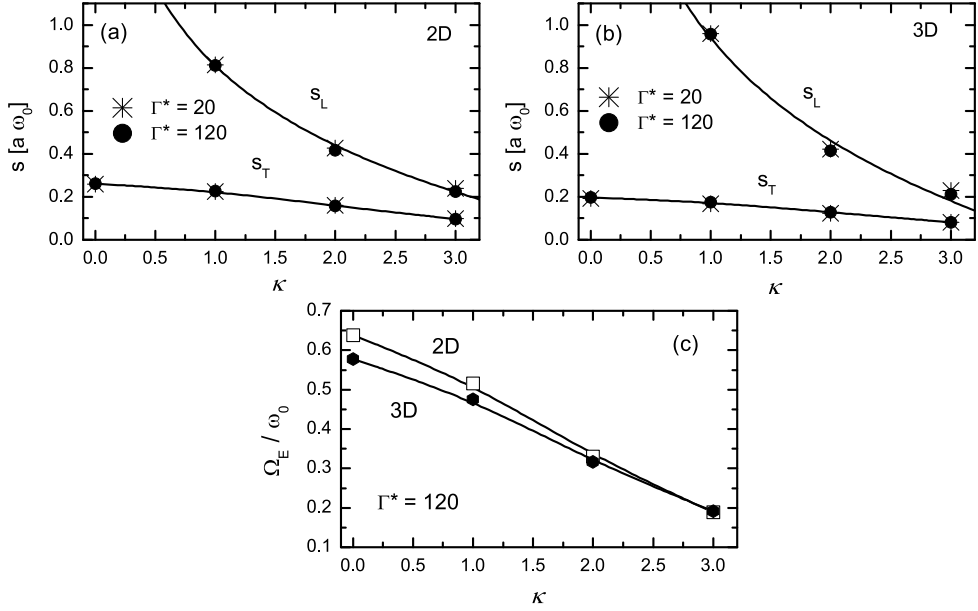}
\end{center}
\caption{2D and 3D Yukawa and Coulomb liquids:
(a,b) QLCA sound velocities and (c) Einstein frequencies.}
\label{fig:tfig03}
\end{figure}

We now can examine the dispersion relations that emerge from
(\ref{eq:qlcadyna}) and (\ref{eq:qlcaeq2}) in conjunction with
(\ref{eq:dieldisp}a) and (\ref{eq:dieldisp}b). 
We will consider both the 2D and 3D cases with the corresponding
results for the dispersion relations displayed in figures \ref{fig:tfig01}
and \ref{fig:tfig02}, respectively. 
Figure \ref{fig:tfig03} will compare the sound
velocities and Einstein frequencies for these two cases.

Turning first to the 2D case, the longitudinal dispersion relation becomes
\begin{eqnarray}\label{eq:qlcaeq4}
\Omega^2_{\rm L}({\bf k}) &=& \Omega^2_0({\bf k}) + D_{\rm L}({\bf k}) =
\Omega^2_0({\bf k})-\frac{n}{m}\int {\rm d}^2rM_{\mu\nu}(r)\left[
{\rm e}^{i{\bf k} \cdot {\bf r}}-1\right]h(r) \\ &=& \omega^2_{\rm 0,2D}
\left\{\frac{\bar{k}^2}{\left(\bar{k}^2+\kappa^2\right)^{1/2}} +
\frac{\bar{k}^2}{2}\int_0^\infty
\Lambda^{\rm 2D}\left(\bar{k}\bar{r},\kappa\bar{r}\right)h(\bar{r}){\rm
d}\bar{r} \right\}, \nonumber
\end{eqnarray}
where $\bar{r}=r/a$, and
\begin{equation}
\Lambda^{\rm 2D}(x,y)=\frac{{\rm e}^{-y}}{x^2}
\left\{\left(1+y+y^2\right)\left[1-J_0(x)\right] +
3\left(1+y+y^2/3\right)J_2(x)\right\}.
\end{equation}
$J_0$ and $J_2$ are Bessel functions of the first kind.

In the mode frequency in (\ref{eq:qlcaeq4}) the RPA solution and the
additional correlational part expressed in terms of the pair
correlation function $h(r)$ are clearly separated. The result can,
however, be transformed into an alternate form, expressed entirely in
terms of the pair distribution function $g(r)=1+h(r)$. By introducing
the extended dynamical matrix $C_{\mu\nu}({\bf k})$:
\begin{eqnarray}\label{eq:qlcaeq5}
\Omega^2_{\rm L}({\bf k}) &=& \frac{n}{m}\int {\rm d}^2rM_{\rm L}(r)\left[
  {\rm e}^{i{\bf k} \cdot {\bf r}}-1\right]g(r) 
  \equiv C_{\rm L}({\bf k}) \\ &=&
  \omega^2_{\rm 0,2D} \frac{\bar{k}^2}{2}\int_0^\infty
  \Lambda^{\rm 2D}\left(\bar{k}\bar{r},\kappa\bar{r}\right)g(\bar{r}){\rm
  d}\bar{r}. \nonumber
\end{eqnarray}
This result shows that the RPA contribution can be interpreted in
terms of the same physical model as the QLCA: in this unified
formulation the RPA force experienced by the oscillating particle is
due to the mean field [$h(r)=0$] only. Moreover, a reflection on the
origin of the ``1'' term in the integrand identifies it as the
generator of the Einstein frequency, the frequency of oscillation of a
single particle in the frozen immobile environment of the other
particles (see more discussion below):
\begin{equation}\label{eq:glcEinst}
\Omega^2_{\rm E} = \frac{n}{m}\int {\rm d}^2rM_{\rm L}(r)g(r) =
\omega^2_{\rm 0,2D} \frac{1}{2}\int_0^\infty \frac{{\rm
d}\bar{r}}{\bar{r}^2} {\rm e}^{-\kappa \bar{r}} \left[1+\kappa \bar{r}+
(\kappa \bar{r})^2 \right]g(\bar{r}).
\end{equation}
The $k\rightarrow 0$ behavior of the longitudinal mode is still
acoustic, but the correlations reduce the acoustic (sound) speed below
its RPA value. For $k\rightarrow \infty$ the
mode frequency approaches the Einstein frequency $\Omega_{\rm E}$. This
limiting behavior is a remarkable feature of strongly coupled Coulomb
and Yukawa liquids \cite{GKW2,GKW3}.

In contrast to the weakly coupled gas described through the RPA, the
strongly coupled liquid supports a shear maintained transverse
mode. This is reflected in the QLCA through the transverse dispersion
relation
\begin{eqnarray}\label{eq:2d_qlcaT}
  \Omega_{\rm T}^2({\bf k}) &=& D_{\rm T}({\bf k}) \\
  &=&\omega^2_{\rm 0,2D}
  \frac{\bar{k}^2}{2}\int_0^\infty
  \Theta^{\rm 2D}\left(\bar{k}\bar{r},\kappa\bar{r}\right)h(\bar{r}){\rm d}\bar{r} 
  \nonumber\\
  &=&\omega^2_{\rm 0,2D}
  \frac{\bar{k}^2}{2}\int_0^\infty
  \Theta^{\rm 2D}\left(\bar{k}\bar{r},\kappa\bar{r}\right)g(\bar{r}){\rm d}\bar{r}
  \nonumber~~~~{\rm 2D},
\end{eqnarray}
where
\begin{equation}
  \Theta^{\rm 2D}(x,y)=2\frac{e^{-y}}{x^2}\left(1+y+y^2\right)\left[1-J_0(x)\right]
  -\Lambda^{\rm 2D}(x,y).
\end{equation}
The last step in (\ref{eq:2d_qlcaT}) follows from a simple algebraic
identity and it reflects the absence of a mean transverse field in the
medium.

The longitudinal and transverse dispersion curves for selected
$\kappa$ and $\Gamma$ values are displayed in \fref{fig:tfig01}. The
$k\rightarrow 0$ behavior of the transverse mode is also acoustic, but
the correlation maintained acoustic speed is substantially below its
longitudinal counterpart. However, the result that the transverse mode
extends all the way to $k=0$ is spurious: the liquid is unable to
support a shear wave in the uniform limit. The reason for this flaw is
well understood: it has to be sought in the neglect of the
migrational-diffusional damping. The introduction of a
phenomenological collision frequency \cite{KRDW00} or of a
semi-phenomenological extension of the QLCA (studied sofar only for
the Coulomb case - see below) \cite{GKW2} provide an acceptable
remedy.

For $k\rightarrow \infty$ the transverse mode also approaches the same
Einstein frequency $\Omega_{\rm E}$ as the longitudinal mode, as
dictated by the isotropy of the liquid.

Also shown are in \fref{fig:tfig03} the $\kappa$ and $\Gamma$
dependences of the acoustic speeds and of the Einstein frequency. For
moderate $\kappa$ values the sound velocities can also be obtained
from the semi-analytic formulae \cite{KHDR_prl}:

\begin{eqnarray}\label{eq:sl}
s_{\rm L}^2 = \frac{\omega_{\rm 0,2D}^2 a^2}{\kappa} \Biggl[ 1 -
\frac{\kappa}{2} \biggl( \frac{5}{8}-\frac{\kappa^2}{2}
\frac{\partial}{\partial \kappa^2} + \frac{3 \kappa^4}{2}
\frac{\partial^2}{\partial \kappa^4} \biggr) \frac{\beta
|E_{\rm c}|}{\Gamma} \Biggr],  \nonumber \\
s_{\rm T}^2 = \frac{\omega_{\rm 0,2D}^2 a^2}{2} \biggl(
\frac{1}{8}-\frac{\kappa^2}{2} \frac{\partial}{\partial \kappa^2} -
\frac{\kappa^4}{2} \frac{\partial^2}{\partial \kappa^4} \biggr)
\frac{\beta |E_{\rm c}|}{\Gamma},
\end{eqnarray}
where $E_{\rm c} = (n/2) \int  \phi(r) [g(r)-1] {\rm d}r^2$ is the
correlation energy per particle.

Turning now to the 3D case, the formal results of the previous
derivation can, {\it mutatis mutandis}, be taken over, with the
understanding that the explicit forms of the RPA frequency
$\Omega_0^2({\bf k})$ and the kernels
$\Lambda\left(\bar{k}\bar{r},\kappa\bar{r}\right)$ and
$\Theta\left(\bar{k}\bar{r},\kappa\bar{r}\right)$ are different
from their 2D counterparts.

\begin{equation}\label{eq:3ddisp1}
\Omega_0^2({\bf k})=\omega_{\rm
  0,3D}^2\frac{\bar{k}^2}{\bar{k}^2+\kappa^2}~~~~~~{\rm 3D}
\end{equation}

\begin{eqnarray}
\Lambda^{\rm 3D}(x,y) &=&  -2 \frac{{\rm e}^{-y}}{x} \Bigg[ \left(
1+y+y^2 \right) \left( \frac{\sin(x)}{x} + 3\frac{\cos(x)}{x^2}
-3\frac{\sin(x)}{x^3} \right)
\nonumber \\
&&- \frac{y^2}{6} \left( 1+ 3 \frac{\sin(x)}{x} +
12\frac{\cos(x)}{x^2} -12\frac{\sin(x)}{x^3}\right) \Bigg]
\end{eqnarray}
and
\begin{equation}\label{eq:3ddisp3}
\Theta^{\rm 3D}(x,y)= \frac{1}{2} \left[ \frac{{\rm e}^{-y}}{x} y^2
\left( 1-\frac{\sin(x)}{x}\right) - \Lambda^{\rm 3D}(x,y) \right].
\end{equation}
There are now 2 degenerate transverse modes.

The expression for the Einstein frequency is also modified:
\begin{equation}\label{eq:3deinstein}
\Omega_{\rm E}^2=\frac{n}{m} \int {\rm d}^3r~M_{\rm L}(r)g(r)= \omega_{\rm
  0,3D}^2\frac{\kappa^2}{3} \int_0^\infty {\rm
  d}\bar{r}~\bar{r}{\rm e}^{-\kappa \bar{r}}g(\bar{r}).~~~~~~{\rm 3D}
\end{equation}

The 3D longitudinal and transverse dispersion curves for selected
$\kappa$ and $\Gamma$ values are displayed in
\fref{fig:tfig02}. The $\kappa$ and $\Gamma$ dependences of the 3D
acoustic speeds and of the Einstein frequency are shown together with
their 2D counterparts in \fref{fig:tfig03}. Finally, for moderate
$\kappa$ values the sound velocities can again be obtained from the
semi-analytic formulae \cite{KRDW00,RosKal97}

\begin{equation}\label{eq:3dsound}
s_{\rm L}^2=\omega_{\rm 0,3D}^2 a^2 \left\{ \frac{1}{\kappa^2}
+ \frac{2}{15} \int_0^\infty \bar{r} {\rm e}^{-\kappa \bar{r}}
\left[ 1+\kappa \bar{r}+\frac{3}{4}(\kappa \bar{r})^2
\right] \left[ g(\bar{r})-1 \right] {\rm d}\bar{r}\right\}
\end{equation}
and 
\begin{equation} 
s_{\rm T}^2=\omega_{\rm 0,3D}^2 a^2 \left\{ -
\frac{1}{15} \int_0^\infty \bar{r} {\rm e}^{-\kappa \bar{r}}
\left[ 1+\kappa \bar{r} - \frac{1}{2}(\kappa \bar{r})^2
\right] \left[ g(\bar{r})-1 \right] {\rm d}\bar{r} \right\}.
\end{equation}

The results of a comparison between the dispersion properties of the 
collective modes in the 2D and 3D systems (assuming the same interparticle
distance) may be summarized as follows.

\begin{enumerate}
\item{While for finite $\kappa$ values the qualitative behaviors of 
the two systems are very much the same, there is the well known 
fundamental difference in the $\kappa = 0$  Coulomb limit between 
the 2D and 3D systems as to the small $k$ dispersion of the longitudinal 
mode: $\omega(k \rightarrow 0) \propto \sqrt{k}$ for 2D, but in the 3D case
$\omega (k=0) = \omega_{\rm 0,3D}$, the 3D plasma frequency.}
\item{Not unrelated to this difference is the behavior of the 
longitudinal acoustic speeds at finite $\kappa$ values: 
since in 2D $s \propto 1/\sqrt{\kappa}$ and 
in 3D $s \propto 1 / \kappa$, for small $\kappa$ the latter exceeds
the former by the factor $\sqrt{3/2\kappa}$.}
\item{In contrast, the transverse acoustic speeds exhibit only a 
mild $\kappa$ dependence and it is the 2D speed that is slightly higher 
than its 3D counterpart.}
\item{A similar 2D dominance prevails for the respective Einstein frequencies
that govern the $k \rightarrow \infty$ behavior of the modes: 
in 2D the Einstein frequency assumes, for any $\kappa$, a somewhat higher  
value than in 3D.}
\end{enumerate}

While (i) and (ii)  are effects originating from the basic difference 
caused by the long range behavior  of the Coulomb potential in a 2D vs. 
a 3D geometry and are already reflected in the RPA description, (iii) and 
(iv) are correlational phenomena and they point at the more important 
role the correlations play in 2D than in 3D.

As a closing comment, it should be re-emphasized that the QLCA ignores
possible damping mechanisms and Doppler shift, phase mixing, etc., due
to the migrational-diffusive motion of the particles and of velocity
dispersion. A method has been recently proposed
\cite{hania3,hania2,hania1,Kiel} for the extension of the QLCA to take
some of the neglected effects into account by combining the $D_{\rm L}(k)$,
$D_{\rm T}(k)$ as local field factors with the Vlasov density-density
response. In this approximation
\begin{equation}
\varepsilon_{\rm L,T}=1-\frac{\varphi(k)\chi_{\rm 0,L,T}({\bf k},\omega)}
{1-\varphi(k)\chi_{\rm 0,L,T}({\bf k},\omega)[D({\bf
k}) / \Omega_0^2({\bf k})]},
\end{equation}
where $\chi_{\rm 0,L,T}({\bf k},\omega)$ is the longitudinal (transverse)
Vlasov density response function of non-interacting particles. The
application of this formalism to Yukawa systems has not been done yet,
but in an early work \cite{GKW2} it was shown that in a 2D Coulomb
system the combined effect of phase mixing and Landau-damping leads to
the elimination of oscillations in the dispersion curve. The effect of
Landau damping, which is not expected to play a major role at high
$\Gamma$ values, is probably overestimated in this work.

\subsection{Lattice phonons}

With the caveat that the present Review addresses primarily the strongly
coupled liquid state, it will be still useful to provide an overview
of the phonon dispersion in a 2D or 3D Yukawa crystal. Such an overview
will help to understand the structure of the liquid state in terms of
a model resembling a disordered lattice and to view the collective
modes in the liquid as being akin to the phonon excitations in the
lattice.

The phonon dispersion is traditionally calculated in terms of the
lattice dynamical matrix defined as
\begin{equation}
C_{\mu\nu}({\bf k})=-\frac{1}{m}\sum_{i,j}M_{\mu\nu}\left( {\bf
r}_i-{\bf r}_j\right) \left[ {\rm e}^{-i{\bf k}\cdot ( {\bf r}_i-{\bf
r}_j)}-1\right],
\end{equation}
with a summation over all the lattice sites $j$, keeping $i$ fixed
$({\bf r}_i=0)$. The resemblance to the extended QLCA dynamical matrix is not
accidental. In contrast to the QLCA equivalent, however, the lattice
dynamical matrix reflects the symmetry of the underlying lattice and
not the rotational invariance of the isotropic
liquid. Nevertheless, a dielectric tensor can be constructed along the
same line in terms of the matrix $D_{\mu,\nu}({\bf k})$, which is
defined now as $C_{\mu,\nu}({\bf k})$ with its mean field
contribution removed
\begin{equation}
D_{\mu\nu}({\bf k})=C_{\mu\nu}({\bf k})+\frac{1}{m}\int{\rm
  d}^Dr~M_{\mu\nu}({\bf r})\left[ {\rm e}^{-i{\bf k}\cdot {\bf r}}-1\right].
\end{equation}
This leads to a structure analogous to (\ref{eq:qlcaeq2}):
\begin{equation}
\varepsilon_{\mu\nu}({\bf
  k},\omega)=\delta_{\mu\nu}-\left[\frac{\Omega_0^2({\bf k})}{{\bf
      1}-{\bf D}({\bf k})} \right]_{\mu\nu}.
\end{equation}
The diagonalization of $\varepsilon_{\mu\nu}$ (or of $C_{\mu\nu}$) is
now possible in the coordinate system of the eigenvectors, whose
orientations, in general, do not coincide either with the direction of
${\bf k}$ or with the crystallographic axes.

To find the eigenmodes one can follow the traditional method (see
  e. g. \cite{Maradudin1,Maradudin2}) of solving the secular equation
\begin{equation}
||\omega^2-C_{\mu\nu}({\bf k})||=0,
\end{equation}
or continue to follow the path of working with the dielectric
tensor. This latter approach ensures that continuity with the liquid
and RPA formalism is maintained. The dispersion relation in terms of
$\varepsilon_{\mu,\nu}$ becomes
\begin{equation}
k_\mu\varepsilon_{\mu\nu}k_\nu=0,
\end{equation}
an obvious generalization of (\ref{eq:dieldisp}a). In fact, except in
the degenerate isotropic case, it also includes the transverse
relation (\ref{eq:dieldisp}b).

\begin{figure}[h!]
\begin{center}
\epsfxsize=13cm
\epsfbox{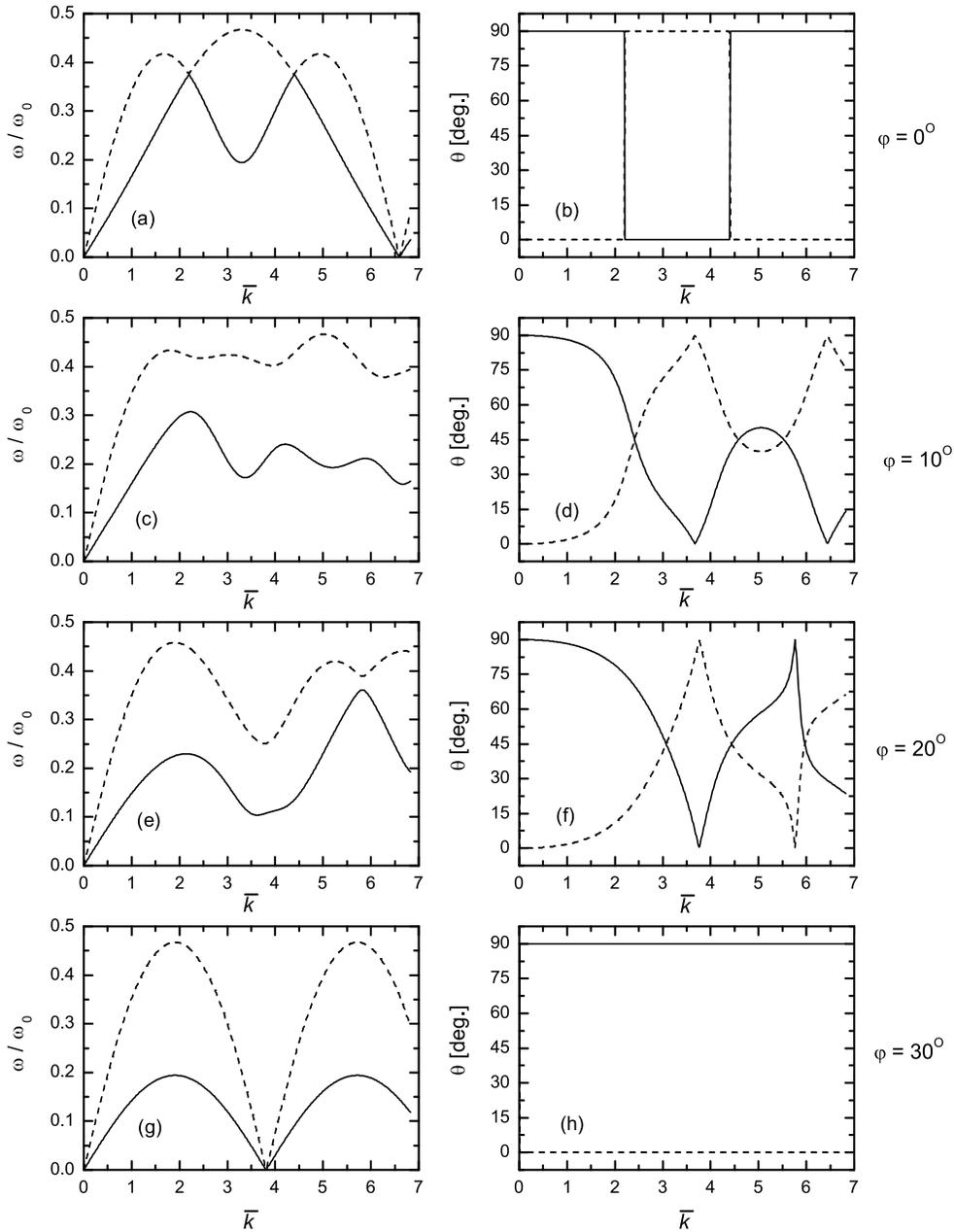}
\end{center}
\caption{2D Yukawa system: lattice dispersion curves and polarizations
($\kappa=2$) for different angles of propagation. $\varphi$ is
measured from the axis pointing towards the nearest neighbor, $\Theta$
is the polarization angle measured with respect to the propagation
vector ${\bf k}$. Partly reproduced from Ref. \cite{Sullivan},
copyright (2006) by Institute of Physics Publishing.}
\label{fig:fig04}
\end{figure}

The 2D Yukawa system crystallizes in a triangular (hexagonal)
lattice. The phonon spectrum was first calculated by Peeters and Wu
\cite{Peeters87}, followed by Wang {\it et al.} \cite{Wang01}; a
definitive calculations of the dispersion and the polarization for all
propagation angles are given in \cite{Sullivan,IEEE07b}. These results
are shown in \fref{fig:fig04}. The mode polarizations are purely
longitudinal or transverse for propagation along the crystallographic
axes ($\varphi = 0^o$ and $30^o$) only, otherwise they are mixed as
shown in the figure. $\varphi$ is the propagation angle measured from
the axis pointing towards the nearest neighbor. The angle $\Theta$
indicated in \fref{fig:fig04} is the polarization angle measured with
respect to the propagation vector ${\bf k}$. The dispersion curves are
periodic in $k$, but the period is simply the reciprocal lattice
constant only along $0^o$ and $30^o$; for intermediate angles it is
much longer, given by the formula
\begin{equation}
\bar{k}_0=\frac{4\pi}{\sqrt{3}}\sqrt{p^2+pq+q^2},
\end{equation}
where $p$ and $q$ are minimal integers satisfying
\begin{equation}
\tan\left(\frac{\pi}{6}-\varphi\right)=\sqrt{\pi}\frac{p}{p+2q}.
\end{equation}

\begin{figure}[h!]
\begin{center}
\epsfxsize=6cm
\epsfbox{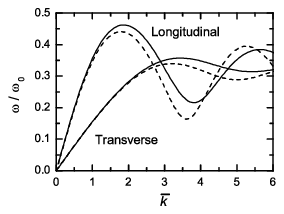}
\end{center}
\caption{2D Yukawa system: angularly averaged lattice (dashed lines)
  and QLCA (solid lines) dispersions of longitudinal and transverse
  modes using pair-correlation [$g(r)$] data from MD
  simulation at $\Gamma=360$ and $\kappa=2$. Reproduced from
  Ref. \cite{Sullivan}, copyright (2006) by Institute of Physics
  Publishing. }
\label{fig:fig05}
\end{figure}

The dispersion curves of the lattice and those of the strongly coupled
liquid do not show much resemblance. Yet, if one views the liquid as
an aggregate of locally ordered domains whose symmetry axes are
randomly distributed, then the similarity to the liquid dispersion
should be sought in a suitably {\it averaged} dispersion of the
lattice. The strong angular dependence of the period $k_0$ suggests
that an angular average should generate through phase mixing a smooth
dispersion. This was carried out by projecting out the
longitudinal and transverse components of the eigenmodes and comparing
their respective angular averages with the longitudinal and transverse
liquid modes \cite{IEEE07b}. \Fref{fig:fig05} shows that the
agreement is quite reassuring. In principle, of course, one has to
distinguish between the spectrum of an average of configurations and
the average of the spectra of each of the configurations: that this
observation notwithstanding the similarity persists can be taken as an
indication that the sizes of the ordered domains in the liquid state
are sufficiently large to diminish the effect of interaction between
the domains.

\begin{figure}[h!]
\begin{center}
\epsfxsize=13cm
\epsfbox{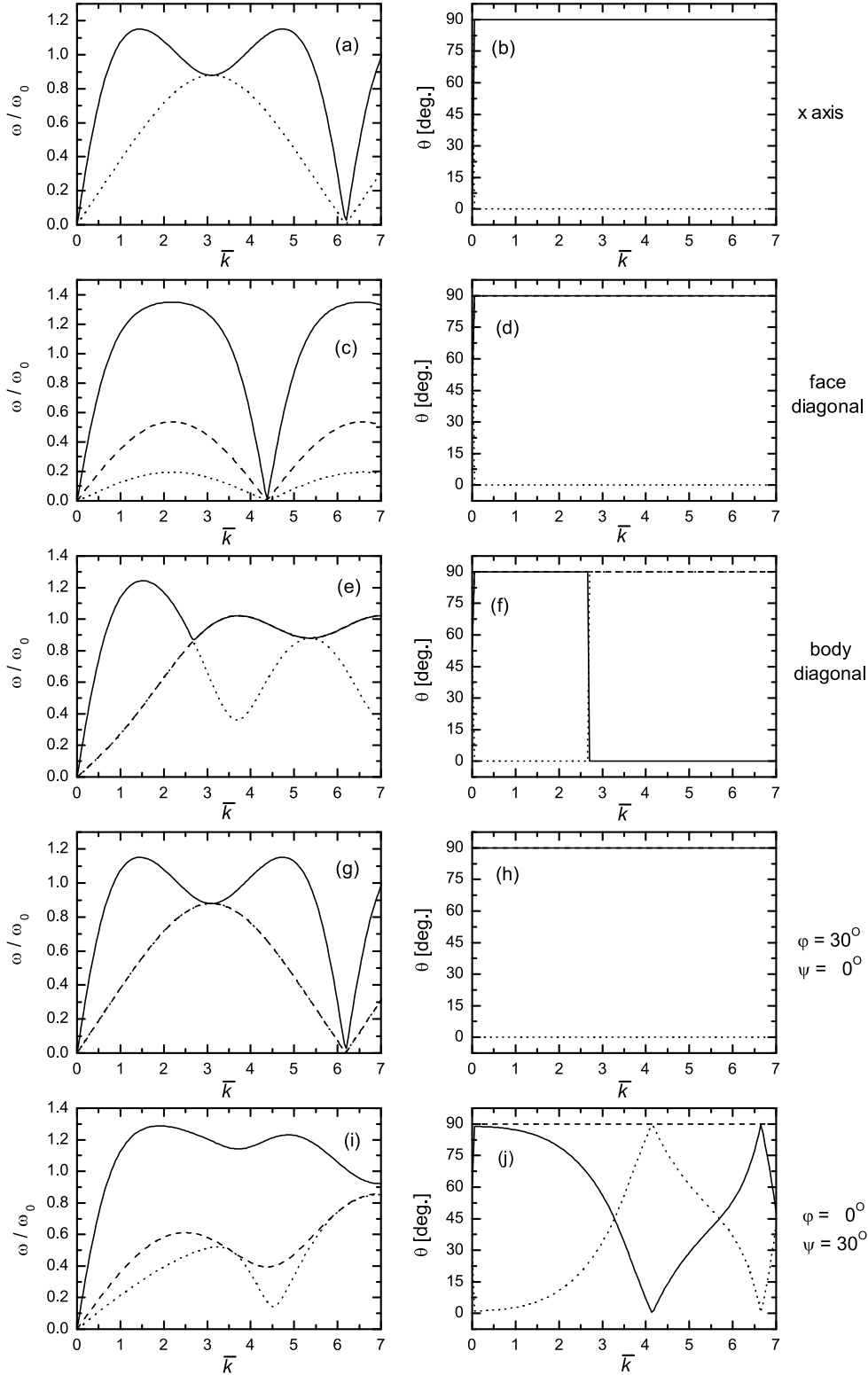}
\end{center}
\caption{3D Yukawa system at $\kappa=1$: bcc lattice dispersion curves
(left column) and polarizations (right column). $\Theta$ is the
polarization angle measured with respect to the propagation vector
${\bf k}$. Continuous line: longitudinal; dashed line: first
transverse; dotted line: second transverse polarizations. Note that
the two transverse polarizations may be, but in general are not
degenerate. $\varphi$ is the polar angle in the $(x,y)$ plane and
$\psi$ is the azimuthal angle measured from the $z$ axis.}
\label{fig:fig06}
\end{figure}

The 3D Yukawa system crystallizes in a bcc or a fcc lattice (depending
on the value of $\kappa$). A phase diagram has been given by Hamaguchi
{\it et al.} \cite{hamaguchi1997}. Due to the existence of 3 rather
than 2 eigenmodes and to their dependence both on the azimuthal and
the polar angles of propagation a much more complex phonon spectrum is
expected than in 2D. So far no systematic published study of this
spectrum seems to exist; in an unpublished work, however, Sullivan,
Kalman and Kyrkos \cite{SullivanUP} have generated a series of
dispersion and polarization diagrams. A sample of
these, for a number of $\varphi$ and $\psi$ angles, including
propagations along the principal crystallographic axes is given in
\fref{fig:fig06}. ($\Theta$ is the polarization angle measured with
respect to the propagation vector ${\bf k}$, $\varphi$ is the polar
angle in the $(x,y)$ plane and $\psi$ is the azimuthal angle measured
from the $z$ axis.) Since no averaging has been performed, their
comparison with the liquid spectrum at the present time is difficult.

\clearpage

\subsection{Einstein frequencies}

In addition to the collective excitations, Einstein frequencies
represent a dynamical manifestation of the strong interaction in
Yukawa systems.  Einstein frequencies, as noted above, are the
frequencies of oscillation of a single particle of the system (the
``test particle'') around its equilibrium position in the immobilized
frozen environment of the other particles of the system. For obvious
reasons, from the experimental point of view the ``freezing'' of the
system but one particle is not a realistic proposition. 

Thus until the advent of dusty plasma experiments Einstein frequencies
were considered more of a theoretical construct than an observable
quantity. The realization, however, that in the strongly coupled
liquid state (but not in the crystalline solid) they represent the
asymptotic $k\rightarrow \infty$ limit of the mode dispersion has
promoted the Einstein frequency to the rank of observable quantities
\cite{QuinnEF,NSG00}.

In the crystalline solid state, where the test particle occupies a
lattice site, the assumption that the potential experienced by the test
particle is a quadratic function of the coordinates with a positive
definite second derivative is in accord with the basic model of the
harmonic theory of phonons. The maximum number of eigenfrequencies of
oscillation is equal to the dimensionality of the system ($\cal{D}=2$ or
$\cal{D}=3$); because of the lattice symmetry induced degeneracy the actual
number may be less than $\cal{D}$. In a disordered lattice the degeneracy is
removed and the frequencies depend on the actual realization of the
disorder. In this case one has to distinguish between the
``microscopic'' Einstein frequencies ($\omega_{\rm E}$) each of which
is generated by a particular realization of the disorder and
characterized by a distribution over the ensemble, and their ensemble
average $\Omega_{\rm E}=\sqrt{\langle \omega_{\rm E}^2\rangle}$. It
is this latter that will be continued to be referred to as ``Einstein
frequency'' in the rest of this paper.

In addition, it is useful to consider the quantity
\begin{equation}
\bar{\omega}_{\rm E}^2=\sum_{i=1}^{\cal{D}}\omega_{{\rm E},i}^2,
\end{equation}
i.e. the sum of the squared eigenfrequencies in a particular
realization. Obviously, $\langle \omega_{\rm E}^2\rangle= \langle
\bar{\omega}_{\rm E}^2\rangle$ but the distributions of $\omega_{\rm
E}^2$ and $\bar{\omega}_{\rm E}^2$ can be quite different.

In a strongly coupled {\it liquid} the very notion of ``equilibrium
position'' is questionable. Nevertheless, the ``quasilocalization''
condition, the basic tenet of the QLCA, is well satisfied for high
$\Gamma$ values, as demonstrated by MD experiments \cite{DKG_caging}
the details of which will be discussed below. It is in this sense that
the notion of the Einstein frequency and its distribution can be
extended to the case of the strongly coupled liquid.

In a 3D Coulomb crystal the Einstein frequency is determined solely by
the background, unaffected by the distribution of the (frozen)
particles. This is the consequence of Gauss Theorem which, in turn,
follows from the Poisson Equation that the 3D Coulomb potential
satisfies. In this case
\begin{equation}\label{eq:we3d}
\omega_{\rm
E}^2=\frac{1}{3\varepsilon_0}\frac{Q^2n}{m}=\frac{1}{3}\omega_{\rm
0,3D}^2.
\end{equation}
In a disordered lattice or in a liquid (\ref{eq:we3d}) is not valid
anymore; it is replaced by the weaker statement
\begin{equation}\label{eq:we3db}
\bar{\omega}_{\rm E}^2=\frac{1}{3}\omega_{\rm 0,3D}^2,
\end{equation}
the so-called Kohn Sum Rule \cite{Brout}, which also follows from the
Poisson Equation.

Thus, in a disordered system while $\omega_E^2$ has a spread,
$\bar{\omega}_E^2$ does not. As to the average, (\ref{eq:we3db}) is
of course also tantamount to
\begin{equation}\label{eq:we3dc}
\Omega_{\rm E}^2=\frac13\omega_{\rm 0,3D}^2.
\end{equation}

For a genuine Yukawa potential the situation is quite different. The
Yukawa potential satisfies the screened Poisson Equation rather than
the Poisson Equation. A useful statement can be made now only for
$\Omega_{\rm E}$, which now can be expressed in terms the average of
the Yukawa potential $\langle \phi \rangle$ as experienced by the test
particle at ${\bf r}=0$ \cite{Bakshi}:
\begin{eqnarray}\label{eq:latt320}
\Omega_{\rm E}^2 &=& \frac{\kappa^2}{3m}\langle\phi\rangle \\
&=&\omega_{\rm 0,3D}^2\frac{\kappa^2}{3}\int_0^\infty{\rm d}\bar{r}~
\bar{r}{\rm e}^{-\kappa r}g(\bar{r}) \nonumber \\
&=&\omega_{\rm 0,3D}^2\frac{1}{3}\left[1+ \kappa^2 
\int_0^\infty{\rm d}\bar{r}~\bar{r}{\rm e}^{-\kappa r}h(\bar{r}) \right]. \nonumber
\end{eqnarray}
(\ref{eq:latt320}) is in agreement with (\ref{eq:3deinstein}), the
result obtained from the QLCA. The third line clearly shows that,
remarkably, in the $\kappa\rightarrow 0$ Coulomb limit the Yukawa
Einstein frequency reduces to the background induced (\ref{eq:we3dc}),
even though the Yukawa system exists without any background. It can
also be noted that $\frac{1}{2}\langle\phi\rangle=E_{\rm int}$ is the
interaction energy density of the system (with [positive] Hartree plus
[negative] correlation contributions). Since the energy is the lowest
in the ordered state, the Einstein frequency must increase with
increasing disorder. According to the known phase diagram of the 3D
Yukawa system \cite{hamaguchi1997} -- as already mentioned -- the
system crystallizes in a bcc or a fcc lattice. The corresponding
Einstein frequencies \cite{SH02}
\begin{eqnarray}
\Omega_{\rm E}^2(\kappa=0)&=& 0.33333~\omega_{\rm 0,3D}^2\\
\Omega_{\rm E}^2(\kappa=1)&=& 0.22293~\omega_{\rm 0,3D}^2\nonumber\\
\Omega_{\rm E}^2(\kappa=2)&=& 0.09416~\omega_{\rm 0,3D}^2\nonumber
\end{eqnarray}
constitute an absolute lower bound.

In the 2D Coulomb system Gauss Theorem does not apply, the background
plays no role and neither the Poisson Equation nor its screened
variant is satisfied. Consequently, the Einstein frequency is
determined by the distribution of the surrounding particles, both for
Yukawa and Coulomb systems. In general
\begin{equation}
\Omega_{\rm E}^2=\frac{1}{m}\langle M_{\mu,\nu}(r=0)\rangle =
\omega_{\rm 0,2D}^2\int_0^\infty\frac{{\rm d}\bar{r}}{\bar{r}^2} {\rm
e}^{-\kappa \bar{r}} \left[\frac{1}{2}(1+\kappa \bar{r})+(\kappa
\bar{r})^2\right] g(\bar{r})
\end{equation}
in agreement with the QLCA result (\ref{eq:glcEinst}).

An argument similar to the one discussed in relation to the 3D case
leads to the conclusion that here also the ordered state exhibits the
lowest Einstein frequency. The lattice structure is now hexagonal, for
which
\begin{eqnarray}
\Omega_{\rm E}^2(\kappa=0)&=& 0.39925~\omega_{\rm 0,2D}^2\\
\Omega_{\rm E}^2(\kappa=1)&=& 0.34433~\omega_{\rm 0,2D}^2\nonumber\\
\Omega_{\rm E}^2(\kappa=2)&=& 0.24347~\omega_{\rm 0,2D}^2\nonumber
\end{eqnarray}
These values constitute then the lowest bound for the 2D Einstein frequencies.

In addition to the frequencies, the Einstein oscillations are also
characterized by their eigenpolarizations. It is the distribution of
the polarization angles which is of interest; this question has been
investigated, however, only for the 2D case \cite{IEEE07b}. In the perfect
hexagonal lattice the degeneracy of the eigenmodes renders this
distribution isotropic. It is also isotropic in the liquid
phase. However, in the intermediate range where the lattice disorder
develops the degeneracy for the microscopic eigenmodes is removed and
the rotational invariance of the distribution is reduced to the
sixfold symmetry of the underlying lattice. More will be shown about
this remarkable effect in Section \ref{sec:twoD}.

\section{Simulation results}\label{sec:simulation}

In this Section we review the results of the extensive MD simulation
work carried out since the beginning of this decade on the dynamical
properties of Yukawa liquids. Most of the work was motivated by the
QLCA theory and accordingly a great portion of the results pertaining
to the areas where QLCA predictions are available are accompanied by
comparisons with the theoretical predictions. However, the information
generated by the simulations goes well beyond those predictions: this
is eminently true for the frequency spectra of the dynamical
density-density and current-current correlation functions [dynamical
structure functions $S(k,\omega)$, $L(k,\omega)$,
$T(k,\omega)$]. Beyond predictions pertaining to the peak positions of
the spectra, identified as the frequencies of the collective
excitations, the QLCA does not provide, apart from some qualitative
estimates, any basis for comparison in this respect. While other
works, based mostly on the memory function formalism
\cite{Barrat1998,Peeters87,Wang01,wake}, have presented theoretical
descriptions of some of the features of the structure functions, we
have made no attempt to relate to these, rather scant, results for the
purpose of comparison with simulations.

\begin{figure}[h!]
\begin{center}
\epsfxsize=13cm
\epsfbox{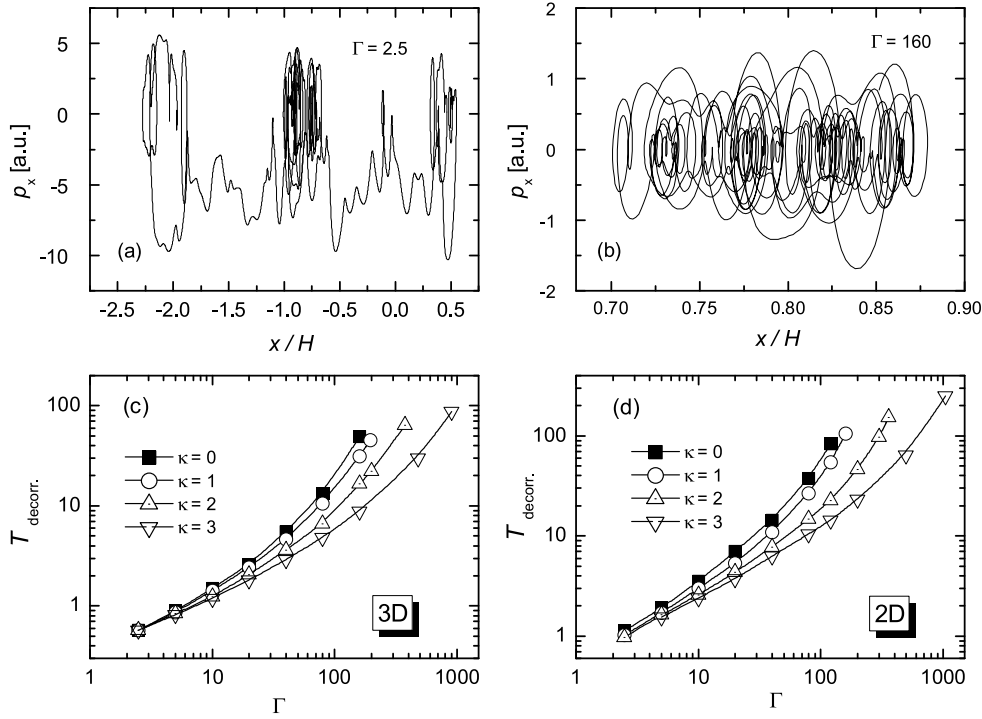}
\end{center}
\caption{\label{fig:decorr}Phase space trajectory segments of a test
  particle in a 3D Coulomb liquid at (a) $\Gamma$ = 2.5 and (b)
  $\Gamma$ = 160. $H$ is the edge length of the simulation
  box. Decorrelation time of the cages ($T_{\rm decorr} = \omega_{\rm
    0} t_{\rm decorr} / 2 \pi$) as a function of $\Gamma$ for the (c)
  3D and (d) 2D systems, for a series of $\kappa$ values, obtained from MD
  simulations. (a,b) Reproduced from Ref.~\cite{DKG_caging}. Copyright (2002) 
  by the American Physical Society. (c,d) 
  Reprinted with permission from [Z. Donk\'o, P. Hartmann, and G. J. Kalman, 
  Phys. Plasmas 10, (5), 1563 (2003)]. Copyright (2003) by the American        
  Institute of Physics, Ref.~\cite{DHK_aps}.}
\end{figure}

As noted in Section \ref{sec:qlca}, the basic hypotheses (i)--(iv)
of the QLCA theory have undergone careful testing by a series of MD
simulation experiments both for Coulomb and Yukawa systems and both
for 2D and 3D configurations \cite{DKG_caging,DHK_aps}. With increasing
$\Gamma$ values, a visual inspection of the potential landscape
clearly indicates the formation of potential wells
\cite{DHK_aps}. Examination of the phase space trajectories reveals a
clear morphological difference between low $\Gamma$ and high $\Gamma$
situations: in the first case the trajectories are open, interrupted
by propagating oscillatory portions, while in the second case the
trajectories are mostly closed and exhibit a loop structure
characteristic of localized oscillatory motion \cite{DKG_caging}. An
example of this behavior is illustrated for a 3D Coulomb liquid in
figures~\ref{fig:decorr}(a) and (b) for $\Gamma$ = 2.5 and $\Gamma$ = 160,
respectively.

The quantification of the relationship between localization and the
strength of the coupling has been carried out by invoking a technique
due to Rabani {\it et al.} \cite{Rabani}. Here a ``cage correlation
function'' was introduced to characterize the gradual disintegration
of the cage of the nearest neighbors and the escape of the caged
particle. The main results shown in figures~\ref{fig:decorr} (c) and
(d) for 3D and 2D Coulomb and Yukawa systems illustrate the duration
(in terms of plasma oscillation cycles) of the caging (decorrelation
time, $T_{\rm decorr}$) as a function of $\Gamma$ and $\kappa$. In the
case of the 3D system, at $\kappa$ = 0 and $\Gamma$ = 160 the cages
decorrelate during $\approx$ 50 plasma oscillation cycles. The
decorrelation time is reduced to a single cycle at $\Gamma \approx$
7. In the case of the 2D system it takes about 100 cycles for the
cages to decorrelate at $\kappa$ = 0 and $\Gamma$ = 120, and we reach
$T_{\rm decorr}$ = 1 at $\Gamma \approx$ 2.5. In the high-$\Gamma$
domain we observe a strong dependence of the decorrelation time on
$\kappa$, both in 3D and 2D systems. At low values of $\Gamma$,
however, $T_{\rm decorr}$ depends only slightly on $\kappa$. The
decrease of the decorrelation time for increasing $\kappa$ can be
compensated by increasing $\Gamma$, as it can be seen in
\fref{fig:decorr}(c) and (d) \cite{DHK_aps}. It is noted that the data
shown in \fref{fig:decorr} convey information about the ``average
behavior'' of the particles, it is however, recognized \cite{DHK_aps}
that the surrounding of individual particles may change in a different
way, due to e.g. avalanche type excitation and migration
\cite{Lai}. Finally we note that the caging of the particles at high
$\Gamma$ values determines many of the systems properties as it has
been discussed by Daligault for 3D Coulomb liquids \cite{Daligault}.

\subsection{Three-dimensional Yukawa liquids}\label{sec:threeD}

The first molecular dynamics simulations on the wave dispersion
relations in the fluid phase of 3D Yukawa systems were reported by
Hamaguchi and Ohta \cite{SH_SCCS,OH_prl}. Their results confirmed the
earlier theoretical predictions of Rosenberg and Kalman
\cite{RosKal97} on the longitudinal wave dispersion and were mostly in
agreement with the simultaneously published full QLCA calculations of
Kalman {\it et al.} \cite{KRDW00}.  They also demonstrated that the
transverse wave dispersion has a cutoff at a long wavelength even in
the case of weak screening.
\begin{figure}
\begin{center}
\epsfxsize=13cm
\epsfbox{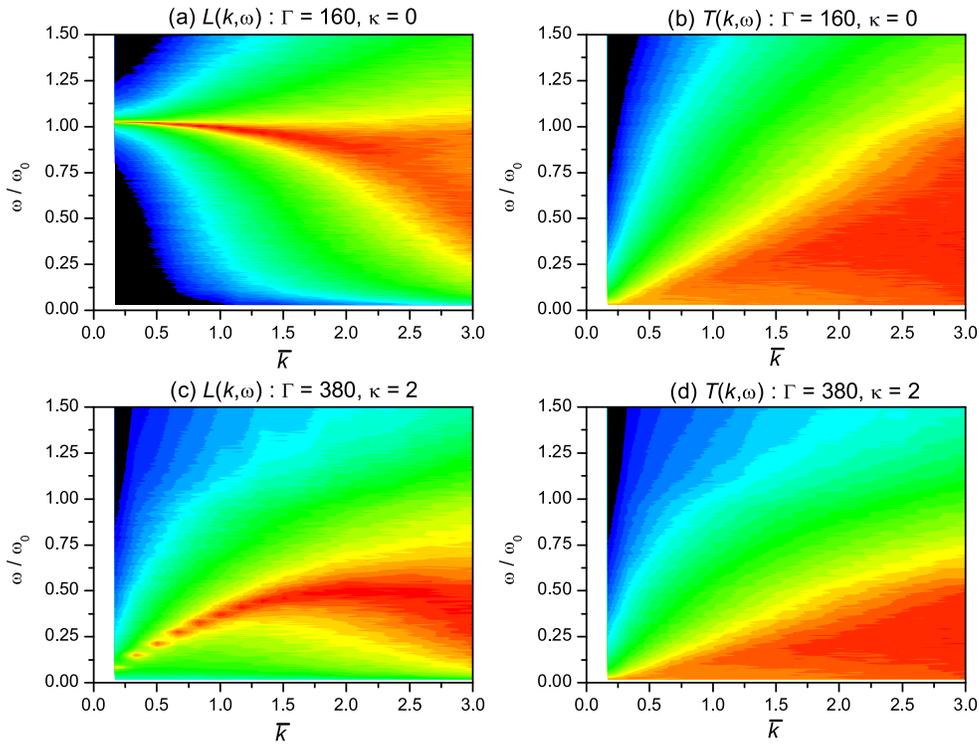}
\end{center}
\caption{\label{fig:spectra1} (Color online) 3D Yukawa and Coulomb liquids:
  spectral decomposition of the longitudinal and transverse current
  fluctuations in Coulomb (a,b) and Yukawa (c,d) plasmas, at $\Gamma$
  = 160, $\kappa$ = 0, and $\Gamma$ = 380, $\kappa$ = 2,
  respectively. (The color coding of the amplitude is logarithmic, it
  only intends to illustrate qualitative features.)}
\end{figure}

This work was followed by a series of MD simulation for the
collective excitations in 3D Yukawa liquids to provide further
comparison with the predictions of the QLCA theory. The simulations --
of which the results are presented here for the first time -- have been
carried out using $N$ = 12~800 -- 15~625 particles.

To illustrate qualitatively the features of the behavior of the
collectives excitations the spectral decomposition of the longitudinal
and transverse current fluctuations is plotted in \fref{fig:spectra1}
for 3-dimensional Coulomb and Yukawa liquids. In the case of the
Coulomb plasma, at low wave numbers the frequency of the longitudinal
($\cal{L}$) mode is concentrated within a narrow frequency range [see
\fref{fig:spectra1}(a)] near the plasma frequency. With increasing
wave number the frequency of the mode gradually spreads over a wider
domain and shows a slightly decreasing tendency. In sharp contrast
with this behavior the $\cal{T}$ mode frequency is spread over a wide
domain, as illustrated in \fref{fig:spectra1}(b). The $\cal{L}$ mode
of the Yukawa system is quite different from that in the Coulomb case,
the wave frequency approaches zero at $\bar{k} \rightarrow 0$ wave
number. The frequency increases with increasing wave number up to
about $\bar{k} = 2.0$, and then starts to decrease slightly. Meanwhile
the frequency distribution gets gradually wider.  The $\cal{T}$ mode
in the Yukawa case appears to be similar to the corresponding mode in
the Coulomb system, although the frequency is lower, due to the weaker
interaction of the particles, as a consequence of the screened
potential.

\begin{figure}
\begin{center}
\epsfxsize=13cm
\epsfbox{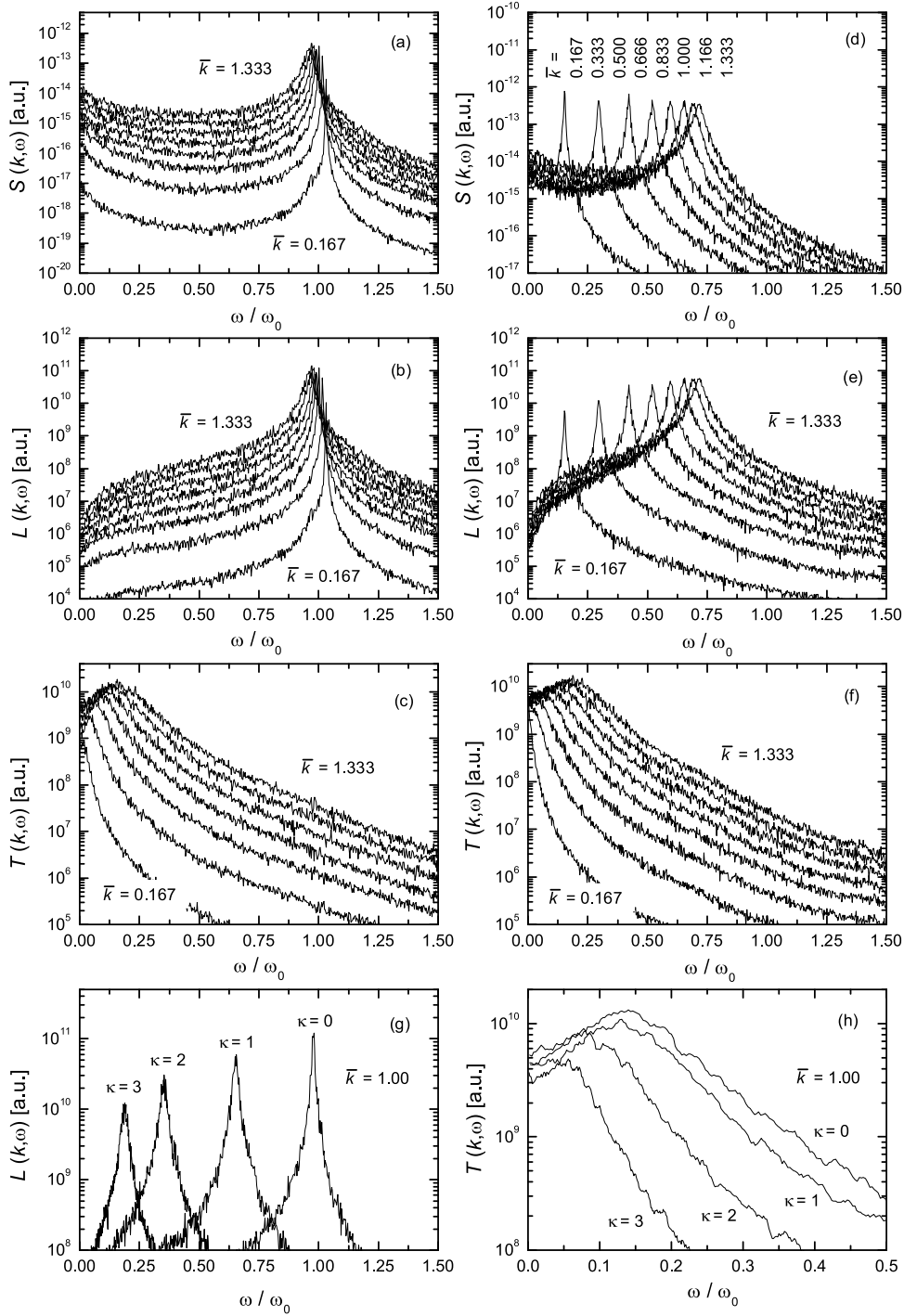}
\end{center}
\caption{\label{fig:y3dspectra} 3D Yukawa and Coulomb liquids: density
  [$S(k,\omega)$] and current [$L(k,\omega)$ and $T(k,\omega)$]
  fluctuation spectra of Coulomb $\Gamma$ = 160,
  $\kappa$ = 0 (a,b,c) and Yukawa $\Gamma$ = 200, $\kappa$ = 1 (d,e,f)
  systems. The curves are plotted for multiples of the smallest
  accessible wave number $\bar{k}_{\rm min}$ = 0.167. (g) and (h) show
  the dependence of $L(k,\omega)$ and $T(k,\omega)$, respectively, on
  $\kappa$ at fixed wave number $\bar k$ = 1.00.  ($\Gamma$ = 360 for
  $\kappa$ = 2, and $\Gamma$ = 1050 for $\kappa$ = 3).}
\end{figure}

\begin{figure}
\begin{center}
\epsfxsize=13cm
\epsfbox{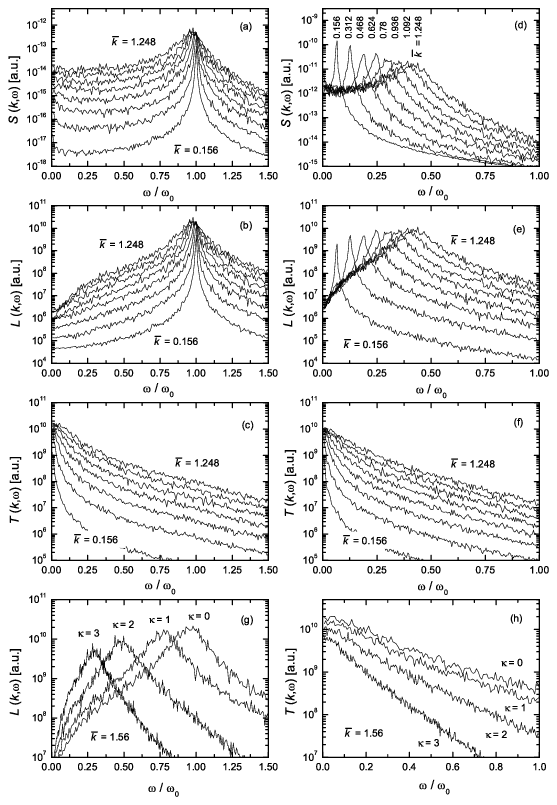}
\end{center}
\caption{\label{fig:y3dspectra-20} 3D Yukawa and Coulomb liquids: density
  [$S(k,\omega)$] and current [$L(k,\omega)$ and $T(k,\omega)$]
  fluctuation spectra of Coulomb $\Gamma$ = 20, $\kappa$
  = 0 (a,b,c) and Yukawa $\Gamma$ = 48, $\kappa$ = 2 (d,e,f)
  systems. The curves are plotted for multiples of the smallest
  accessible wave number $\bar{k}_{\rm min}$ = 0.156. (g) and (h) show
  the dependence of $L(k,\omega)$ and $T(k,\omega)$, respectively, on
  $\kappa$, at fixed wave number $\bar k$ = 1.56.  ($\Gamma$ = 25 for
  $\kappa$ = 1, and $\Gamma$ = 114 for $\kappa$ = 3). }
\end{figure}

\begin{figure}
\begin{center}
\epsfxsize=13cm
\epsfbox{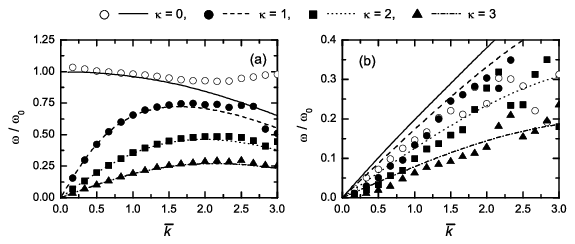}
\end{center}
\caption{\label{fig:y3ddisp}3D Yukawa and Coulomb liquids: dispersion relations
  for the (a) longitudinal and (b) transverse modes.  
  $\opencircle$ : $\Gamma$ = 160, $\kappa$ = 0
  (Coulomb case), $\fullcircle$ : $\Gamma$ = 200, $\kappa$ = 1,
  $\fullsquare$ : $\Gamma$ = 380, $\kappa$ = 2, and $\blacktriangle$ :
  $\Gamma$ = 910, $\kappa$ = 3. Symbols represent molecular dynamics
  results, while the lines correspond to the predictions of the QLCA
  theory.}
\end{figure}

For a better quantitative analysis representative dynamical structure
functions (density fluctuation spectra) $S(k,\omega)$ and spectra of
the longitudinal and transverse current fluctuations, $L(k,\omega)$
and $T(k,\omega)$, are plotted in figures~\ref{fig:y3dspectra} and
\ref{fig:y3dspectra-20}, respectively, for a high-$\Gamma$ and a
medium-$\Gamma$ case. The $S(k,\omega)$ obtained for the Coulomb case
($\Gamma$ = 160, $\kappa$ = 0, see \fref{fig:y3dspectra}(a)) peaks at
nearly the same frequency for the different values of the wave numbers
plotted, which are multiples of $\bar{k}_{\rm min} = 0.167$
(determined by the size of the simulation box). In the presence of
screening (Yukawa potential), as shown in \fref{fig:y3dspectra}(d),
the behavior of $S(k,\omega)$ changes significantly: at $\bar{k}
\rightarrow$ 0 the wave frequency $\omega/\omega_0 \rightarrow$ 0
[$\omega_0$ is defined by (\ref{eq:omegap3d})]. The contrast between
the $\kappa$ = 0 and the $\kappa >$ 0 cases is also well seen in
\fref{fig:y3ddisp}(a), where the dispersion curves derived from the
fluctuation spectra are displayed. The dispersion curves for $\kappa
>$ 0 are quasi-acoustic ($\omega/\omega_0 \propto \bar{k}^{1/2}$),
with a linear portion near $k$ = 0, which gradually extends when
$\kappa$ is increased. The ($\Gamma$,$\kappa$) pairs for which the
dispersion graphs are plotted in \fref{fig:y3ddisp} have been selected
to represent a constant effective coupling $\Gamma^\ast$ = 160. This
definition of $\Gamma^\ast$ relies on the constancy of the first peak
amplitude of the pair correlation function $g(\bar{r})$, similarly to
the case of 2D Yukawa liquids \cite{H2005}.

Peaks in the spectra of the compressional $\cal{L}$ mode [plotted in
  panels (b) and (e) of figures \ref{fig:y3dspectra} and
  \ref{fig:y3dspectra-20}] appear at the same frequency as those in
  the corresponding $S(k,\omega)$ functions, as these functions are
  linked via the relation
\begin{equation}
L(k,\omega) = \frac{\omega^2}{k^2} S(k,\omega). \label{eq:lvss}
\end{equation}

Compared to those characterizing the $\cal{L}$ mode, peaks in the
$\cal{T}$ mode spectra are rather broad, as it can be seen in
panels (c) and (f) of figures \ref{fig:y3dspectra} and
  \ref{fig:y3dspectra-20}. In the case of this mode there is no
dramatic change between the behavior when $\kappa$ changes from zero
to a nonzero value, only the mode frequency decreases, as can be
observed in \fref{fig:y3ddisp}(b).

\begin{figure}
\begin{center}
\epsfxsize=13cm
\epsfbox{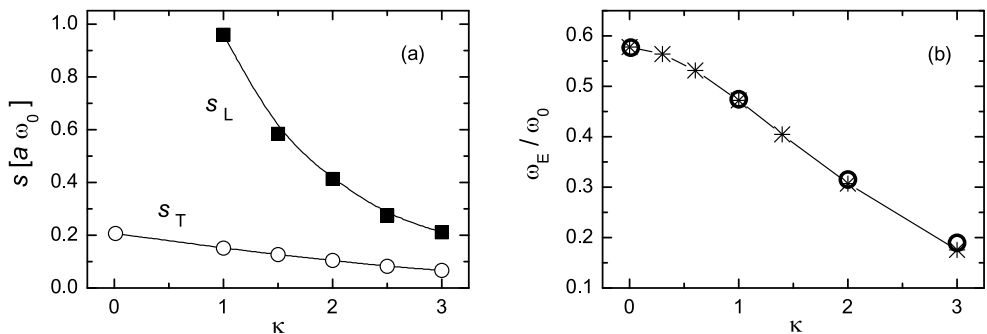}
\end{center}
\caption{\label{fig:y3dsound} 3D Yukawa and Coulomb 
liquids: (a) sound velocities
and (b) Einstein frequency as derived from the QLCA theory (circles)
and Einstein frequencies of fcc lattice (stars) \cite{OH2000}.}
\end{figure}

Comparison of the dispersion relations obtained from the MD
simulations [via $S(k,\omega)$] and QLCA calculations [see equations
(\ref{eq:3ddisp1})--(\ref{eq:3ddisp3})] is presented in
\fref{fig:y3ddisp}. Here, in the calculations of the QLCA results, we
have made use of the $g(r)$ functions obtained from the MD
simulation. The agreement between the two sets of data is excellent
for the $\cal{L}$ mode, while some difference in the frequency of the
$\cal{T}$ waves can be seen in \fref{fig:y3ddisp}(b). This latter may
originate from the inaccurate determination of the peak positions of
the rather broad $T(k,\omega)$ spectra. It should be noted though that
while the theoretical calculations provide an oscillatory dispersion
curve for $\bar{k} > 3$ (see \fref{fig:tfig02}), simulations provide
reliable results (for collective excitations) for typical liquid-phase
conditions for $\bar{k} \lesssim 3$. (At higher $\bar{k}$ values the
thermal contribution in $S(k,\omega)$ apparently masks the collective
mode peak.). The simulation results here resemble the measured 2D
dispersion curves in the liquid phase \cite{Nunomura05}. Another
difference is the cutoff of the $\cal{T}$ mode dispersion curve at
finite wave numbers. This disappearance of the shear modes for
$\bar{k} \rightarrow 0$ is a well known feature of the liquid state
\cite{HMP75,TK80,SZRT97}, and the sharp cut-off $\omega \rightarrow 0$
for a finite $k$ has also been observed in simulations of Yukawa
systems \cite{Murillo3,OH_prl}. It has been already noted that this
cutoff is not accounted for by the QLCA, as it does not include
damping effects.

The sound velocities, derived in (\ref{eq:3dsound}), are plotted in
\fref{fig:y3dsound}(a), while \fref{fig:y3dsound}(b) displays the
Einstein frequency, which is defined in (\ref{eq:3deinstein}). In the
$\kappa \rightarrow 0$ limit (\ref{eq:3deinstein}) gives $\omega_{\rm
E} = \omega_0 / \sqrt{3}$, and $\omega_{\rm E}$ decreases with
increasing $\kappa$. For comparison, the Einstein frequency data of
Ohta and Hamaguchi for a fcc lattice \cite{OH2000} are also plotted in
\fref{fig:y3dsound}(b). We find an excellent agreement between the two
sets of data.

\begin{figure}[h!]
\begin{center}
\epsfxsize=12cm 
\epsfbox{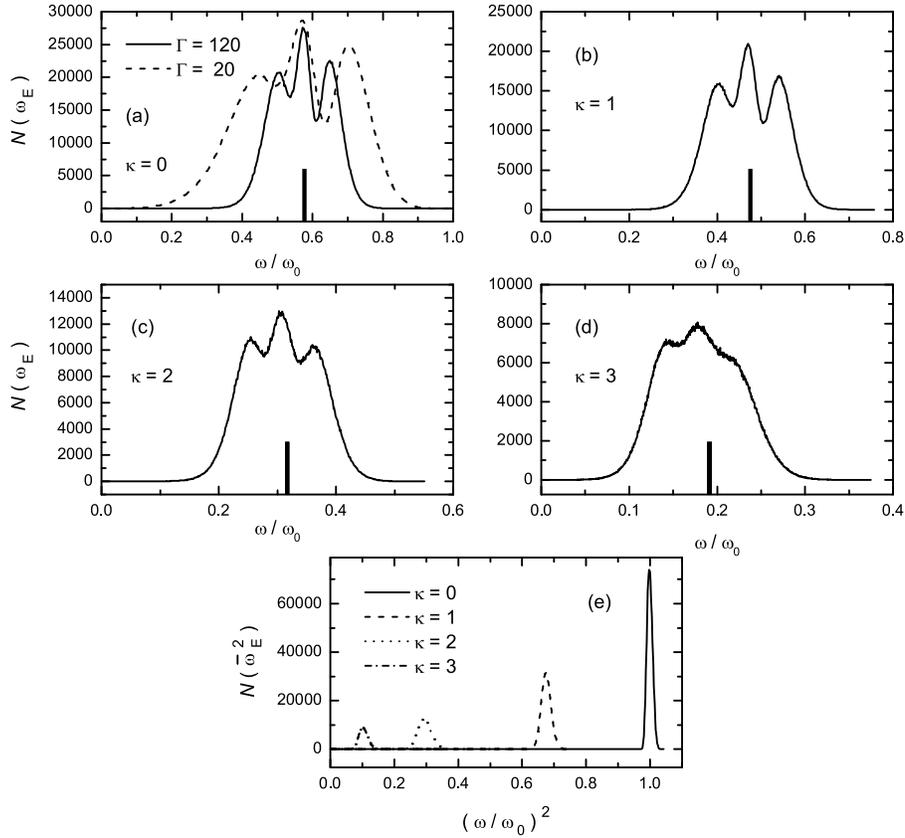}
\end{center}
\caption{\label{fig:3dMDeindisp} 3D Yukawa and Coulomb liquids: 
Einstein frequency
distributions for $\Gamma^\star = 120$: (a) $\Gamma = 120$,
$\kappa=0$; (b) $\Gamma = 150$, $\kappa=1$; (c) $\Gamma = 300$,
$\kappa=2$; (d) $\Gamma = 725$, $\kappa=3$; the vertical bars indicate
values obtained by the QLCA theory. (a) also shows the distribution of
frequencies at the lower coupling value $\Gamma = 20$.  (e)
Distribution of $\bar{\omega}_{\rm E}^2$ for the same systems.}
\end{figure}

Numerical experiments were also performed to determine the
distribution of the microscopic Einstein frequency $\omega_{\rm
E}$. To accomplish this, frequency histograms based on a few hundred,
temporally uncorrelated particle configurations have been constructed.
For the raw (particle position) data the harmonic matrix for every
particle has to be generated:
\begin{equation}
H_{\alpha\beta}^{(i)}=\sum_{j\neq i}^N \frac{\partial^2
  \phi(|{\bf r}_i^{\rm eq}-{\bf r}_j|)}{\partial r_{i,\alpha} \partial r_{i,\beta}},
\label{eq:harmonic}
\end{equation}
where ${\bf r}_i^{\rm eq}$ is the equilibrium position of the $i$-th
particle (local minimum of the potential surface), $\phi(r)$ is the
interaction potential, $\alpha$ and $\beta$ represent the Cartesian
coordinates. The eigenvalues of $H_{\alpha\beta} / m$ are the squared
Einstein frequencies (3 for every particle), while the eigenvectors
provide the polarization of the oscillation.

A series of frequency histograms for an effective coupling parameter 
$\Gamma^\star=120$ and different values of $\kappa$ are 
shown in figures~\ref{fig:3dMDeindisp}(a)-(d). The frequency distributions 
exhibit three peaks (although this is less visible 
in the $\kappa=3$ case). With increasing screening the distribution of
frequencies becomes wider and its mean value is shifted towards lower
frequency. The QLCA results for the Einstein frequency 
[obtained from Eq.(\ref{eq:3deinstein})
using pair correlation functions generated in the MD simulation], 
corresponding to the different values
of $\kappa$ are also indicated in figures~\ref{fig:3dMDeindisp}(a)-(d).
The values are in good agreement with the simulation results. 
The effect of $\Gamma$ at fixed ($\kappa=0$) screening is 
illustrated in \fref{fig:3dMDeindisp}(a). A six time decrease of the
coupling parameter results in approximately doubled width of the
Einstein frequency distribution. 

\Fref{fig:3dMDeindisp}(e) shows the histograms for $\bar{\omega}_{\rm
E}^2$, sums of the 3 microscopic squared Einstein frequencies, for
different values of $\kappa$: there is a qualitative difference
between the $\kappa=0$ Coulomb case where there is only a single
frequency (a narrow peak) and the $\kappa>0$ cases, where a
distribution of frequencies is apparent. The reason for this difference
has been discussed in Section \ref{sec:theory}.

Further information on the collective behavior is contained in the
velocity autocorrelation function (VACF)
\begin{equation}
\label{eq:vacf} Z(t) =  \frac{\langle {\bf v}(t) {\bf v}(0)\rangle }
{\langle |{\bf v}(0)|^2 \rangle},
\end{equation}
where the average is taken over the $N$ particles and different
initial times. 

\begin{figure}[h!]
\begin{center}
\epsfxsize=13cm
\epsfbox{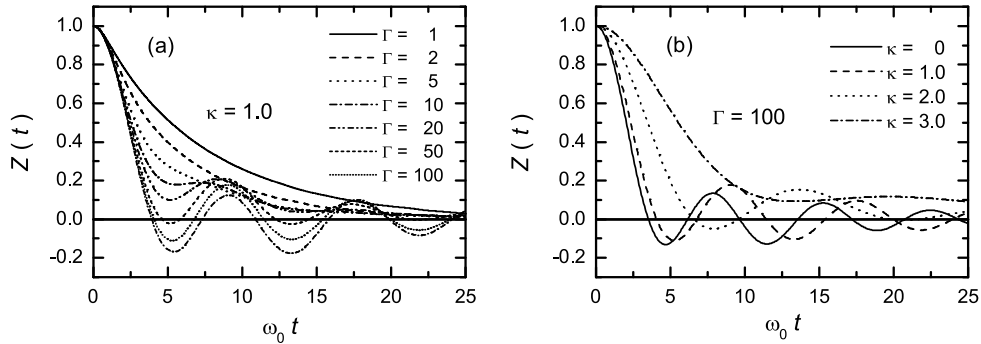}
\end{center}
\caption{\label{fig:y3dvac0}3D Yukawa and Coulomb liquids: (a) velocity
autocorrelation functions at $\kappa$ = 1.0 and a series of $\Gamma$
values; (b) at constant $\Gamma$ = 100, for a series of $\kappa$
values.}
\end{figure}

The behavior of the velocity autocorrelation functions of 3D Yukawa
liquids obtained at several values of the $\Gamma$ and $\kappa$
parameters is illustrated in \fref{fig:y3dvac0}. Analyzing the
behavior of $Z(t)$ at constant $\kappa$, we find a transition from
monotonically decreasing $Z(t)$ into an oscillating type when $\Gamma$
is increased [\fref{fig:y3dvac0}(a)]. Similarly, the shape of $Z(t)$
changes drastically when $\kappa$ is varied at constant $\Gamma$, as
shown in \fref{fig:y3dvac0}(b). For more detailed analysis see
\cite{OH2000}.

\begin{figure}[h!]
\begin{center}
\epsfxsize=13cm
\epsfbox{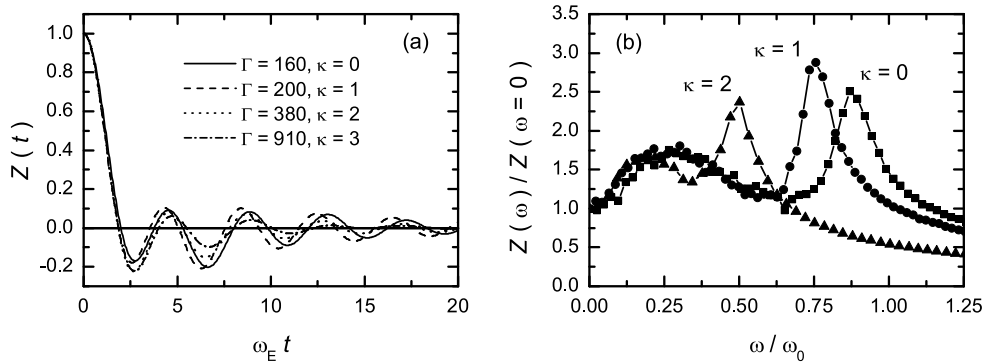}
\end{center}
\caption{\label{fig:y3dvac}3D Yukawa and Coulomb liquids: (a) velocity
autocorrelation functions for a series of $\kappa$ values; the time is
normalized by the Einstein frequency $\omega_{\rm E}$. (b)
corresponding Fourier transforms $Z(\omega)$, for ($\Gamma,\kappa$)
pairs as indicated in (a).}
\end{figure}

Using the Einstein frequency $\omega_{\rm E}$ for the normalization of
time, instead of the plasma frequency $\omega_0$ (as in
\fref{fig:y3dvac0}) the $Z(t)$ functions belonging to the same
$\Gamma^\star$ = 160 for a series of $\kappa$ values are
displayed in \fref{fig:y3dvac}(a). Using this normalization of the
timescale the $Z(t)$ curves exhibit a nearly universal behavior,
where at least the first few peaks of $Z(t)$ nearly overlap. This
observation emphasizes the importance of the Einstein frequency in
the dynamical behavior of the system.

\begin{figure}
\begin{center}
\epsfxsize=13cm 
\epsfbox{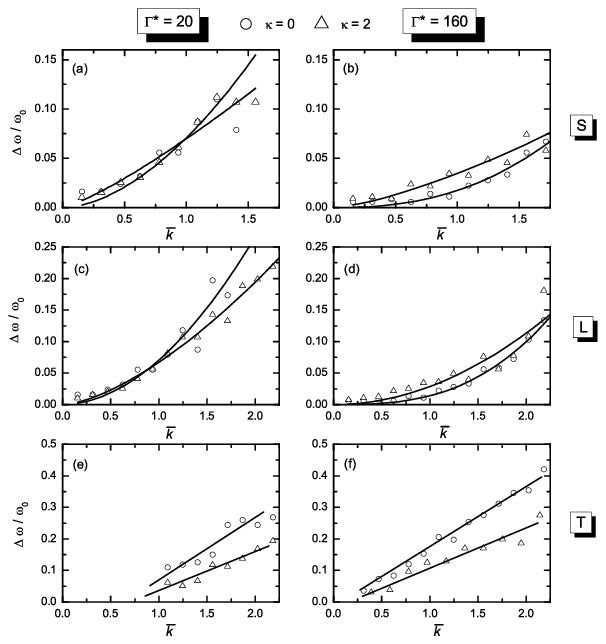}
\end{center}
\caption{\label{fig:3dspecwidth}3D Yukawa and Coulomb liquids: 
spectral line width
of functions $S(k,\omega)$ (a,b), $L(k,\omega)$ (c,d) and
$T(k,\omega)$ (e,f) for $\Gamma^\star=20$ and $\Gamma^\star=160$
effective coupling at $\kappa=0$ and $\kappa=2$.}
\end{figure}

The marked oscillations of the $Z(t)$ function in the Coulomb case is
indeed expected on the basis of the possible coupling between the
single particle motion and long-wavelength plasmons, whose frequencies
are almost independent of $k$. For $\kappa > 0$, however, even though
$\omega(k\rightarrow 0) \propto k$, the oscillations persist; this can
be explained by the fact that the $\omega(k)$ dispersion curve
flattens at higher wave numbers, a corresponding peak in the frequency
distribution develops.

The Fourier transforms $Z(\omega)$ of the VACF functions (obtained at
different $\kappa$ values but for constant effective coupling
$\Gamma^\star$ = 160) are portrayed in \fref{fig:y3dvac}(b). The
dominant peaks in the spectra -- shifting towards lower frequencies
with increasing $\kappa$ -- correspond to the high frequency
oscillations of the $Z(t)$ functions (easily observed visually). As
discussed in previous studies (see e.g. \cite{OH2000,SZRT97}) these
peaks are related to longitudinal current fluctuations, while the
broad features at low frequencies are connected to the transverse
current fluctuations and are related to diffusion properties of the
system. Taking the case of $\kappa$ = 2 as an example,
\fref{fig:spectra1}(c) indicates that most of the energy of the
$\cal{L}$ mode is concentrated around frequencies $\omega / \omega_0
\cong$ 0.5, in correspondence with the peak of the $Z(\omega)$
function shown in \fref{fig:y3dvac}(b). The $T(\bar{k},\omega)$
spectra [see \fref{fig:spectra1}(c)] for any $\bar{k}$ are broader,
compared to the $L(\bar{k},\omega)$ spectra: the fluctuations in the
transverse currents are distributed over a rather broad frequency
domain, again in agreement with the behavior of the corresponding
$Z(\omega)$ function. The observed features of $Z(\omega)$ indicate an
appreciable coupling between single particle motion and collective
excitations in the 2D system.

In addition to the peaks in the frequency spectrum of the dynamical
structure functions a wealth of further physical information is
contained in the detailed structures of these quantities. Of great
importance would be the understanding of the evolution of the width of
the frequency spectra as functions of $k$ and $\Gamma$, since it is
related to the damping of the collective modes. To illustrate the
behavior of the dynamical structure function \fref{fig:3dspecwidth}
shows the widths of the collective mode peaks as a function of wave
number, for effective coupling values $\Gamma^\star$ = 20 and 120, for
$\kappa$ = 0 and 2. It is noted that Murillo has provided a formula
for the width of the transverse current spectrum \cite{Murillo3}.

\clearpage

\subsection{Two-Dimensional Yukawa liquids} \label{sec:twoD}

Most of the available
experimental evidence on waves in complex plasmas relates to 2D
systems (see section \ref{sec:experiment}); 
much less information can be culled from observations on 3D
systems. Thus the understanding of the collective mode structure in
the different phases of the 2D Yukawa system has been of great current
interest: over the past few years a substantial amount of simulation
work has been performed on 2D Yukawa liquids
\cite{KHDR_prl,DHKR_cpp}. In the following we present these MD
simulation results on the dispersion properties of the liquid state
and compare them with the theoretical predictions of the QLCA analysis
of the collective modes.

\begin{figure}
\begin{center}
\epsfxsize=13cm
\epsfbox{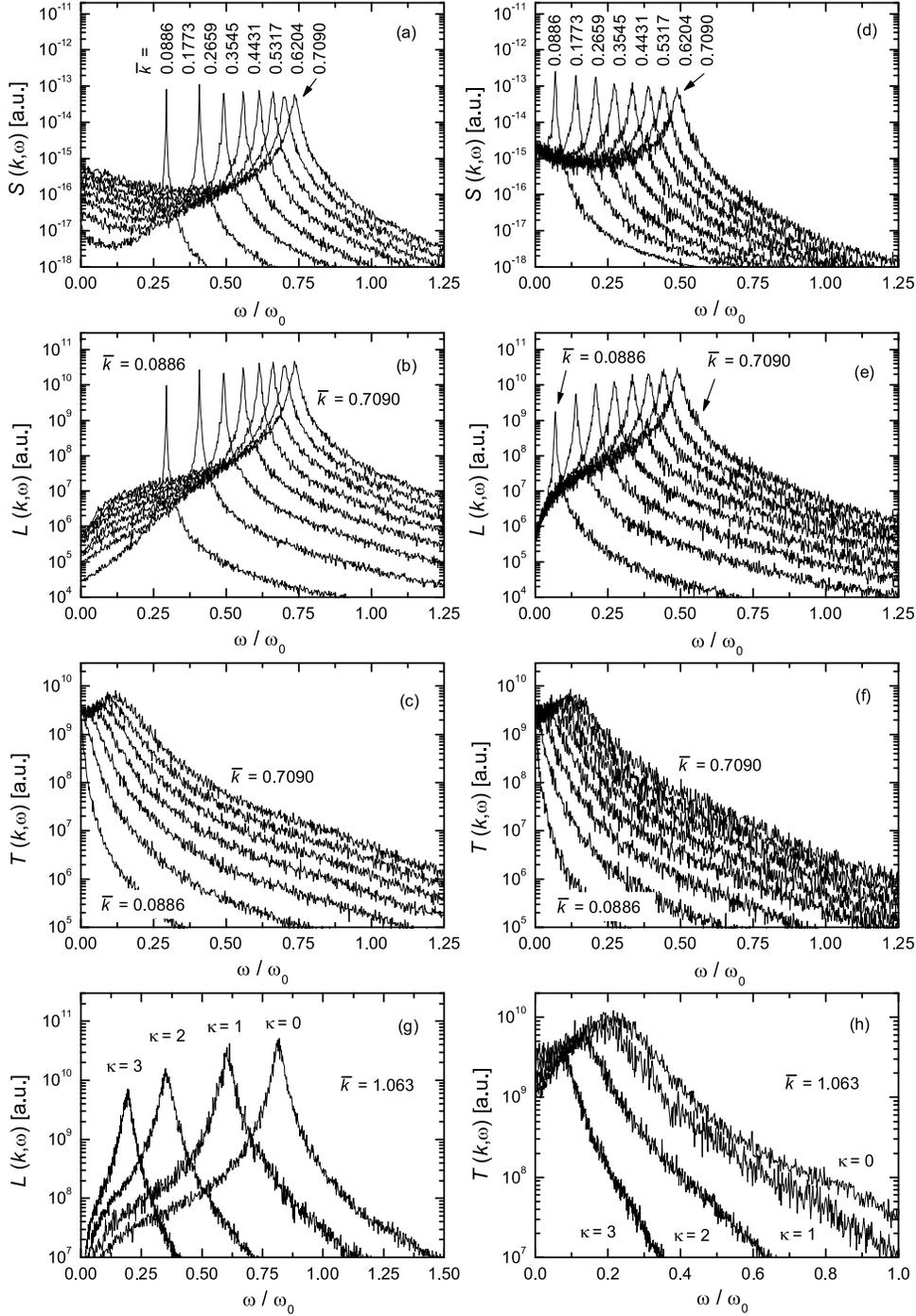}
\end{center}
\caption{\label{fig:y2dspectra}2D Yukawa and Coulomb liquids: density
[$S(k,\omega)$] and current [$L(k,\omega)$ and $T(k,\omega)$]
fluctuation spectra of Coulomb $\Gamma$ = 120, $\kappa$ = 0 (a,b,c)
and Yukawa $\Gamma$ = 160, $\kappa$ = 1 (d,e,f) systems. The curves
are plotted for multiples of the smallest accessible wave number
$\bar{k}_{\rm min}$ = 0.0886. (g) and (h) show the dependence of
$L(k,\omega)$ and $T(k,\omega)$, respectively, on $\kappa$, at fixed
wave number $\bar k$ = 1.063.  ($\Gamma$ = 360 for $\kappa$ = 2, and
$\Gamma$ = 1050 for $\kappa$ = 3).}
\end{figure}

\begin{figure}
\begin{center}
\epsfxsize=13cm
\epsfbox{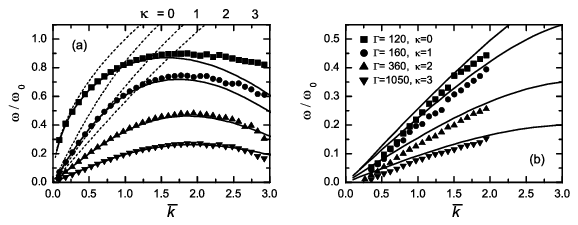}
\end{center}
\caption{\label{fig:y2ddisp}2D Yukawa and Coulomb liquids: dispersion curves for (a)
  the longitudinal ($\cal{L}$) and (b) transverse ($\cal{T}$) modes at
  $\Gamma^{\star}$ = 120 and $\kappa$ = 0, 1, 2, 3. Continuous
  curves: QLCA calculations; symbols: MD simulation; dashed lines: RPA
  dispersions.  Reproduced from Ref. \cite{KHDR_prl}. Copyright (2004)
  by the American Physical Society.}
\end{figure}

Representative density fluctuation spectra, as well as longitudinal
and transverse current fluctuation spectra of the 2D Yukawa liquid are
displayed in \fref{fig:y2dspectra}. The dispersion curves derived from
the simulation spectra $S(k,\omega)$ for both modes are displayed in
\fref{fig:y2ddisp}. The results shown in the latter figure at $\kappa
= 0$ reproduce the known 2D Coulomb dispersion \cite{TK80,GKW1}. With
increasing $\kappa$ the mode frequencies rapidly diminish and the
dispersion deviates more substantially from its RPA value. In the $k
\rightarrow 0$ limit both modes exhibit an acoustic behavior, with
longitudinal and transverse sound velocities $s_{\rm L}$ and $s_{\rm
T}$ \cite{GKW2,GKW1}, see Eq. (\ref{eq:sl}). For the longitudinal mode
the simulation data well corroborate the theoretical predictions with
the proviso already noted in relation to the 3D liquid: that while the
theoretical calculations provide an oscillatory dispersion curve for
$\bar{k} > 3$ (see \fref{fig:tfig01}), simulations provide reliable
results (for collective excitations) for the given conditions for
$\bar{k} \lesssim 3$. In the case of the transverse mode, the
agreement between theory and MD data for moderately high $k$ values is
fairly good; for $k \rightarrow 0$ the agreement is marred by the
QLCA's inability to account for diffusional and other damping effects
\cite{GKW1} that preclude the existence of long wavelength shear waves
in the liquid state. As a result of this damping, a cutoff at a finite
$k_{\rm c}$ and zero frequency develops (a similar phenomenon was
observed in the 3D case \cite{Murillo3,OH_prl}). The $k_{\rm c}$ value
is related to the diffusional-migrational time \cite{GKW1} through
$\tau_{\rm DM} = 1 / k_{\rm c} s_{\rm T}$, where $s_{\rm T}$ is the
transverse sound velocity. Incorporating $\tau_{\rm DM}$, calculated
with the aid of the theoretically predicted $s_{\rm T}$ values, in the
QLCA equations as a phenomenological damping $\nu = 1 / \tau_{\rm DM}$
(by the $\omega \rightarrow \omega + i\nu$ replacement), good
agreement between the theory and the MD data was restored
\cite{KHDR_prl}. The simulations show that the longitudinal mode is
not affected by this damping mechanism: this may indicate that its
characteristic damping time is substantially longer.

\begin{figure}
\begin{center}
\epsfxsize=13cm
\epsfbox{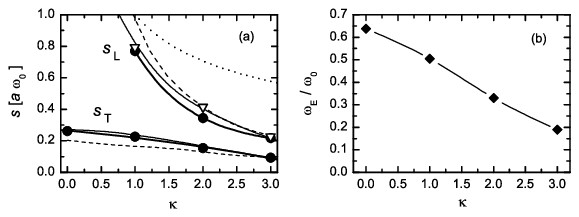}
\end{center}
\caption{\label{fig:y2dspeed} 2D Yukawa and Coulomb liquids: 
(a) sound velocities (heavy
lines with circles: calculated from QLCA at $\Gamma^\star$ = 120, thin
lines: hexagonal crystal lattice \cite{Peeters87}, dotted line: RPA
values for $s_{\rm L}$, and dashed lines: 3D values at $\Gamma$ = 160,
open triangles: thermodynamic sound velocity) \cite{KRDW00,KHDR_prl};
(b) calculated Einstein frequency as obtained from the QLCA formula
for $\Gamma^\star$ = 120. Reproduced from
Ref. \cite{KHDR_prl}. Copyright (2004) by the American Physical
Society.}
\end{figure}

The sound velocities and the Einstein frequency -- given as the $k
\rightarrow \infty$ limit of Eqs. (\ref{eq:qlcaeq4}) or
(\ref{eq:qlcaeq5}) -- are shown in \fref{fig:y2dspeed}. For comparison,
also displayed is the thermodynamic sound velocity generated from the
equation of state of a Yukawa liquid \cite{H2005}. The sound velocities
obtained here are extremely close to those of the hexagonal crystal
\cite{Peeters87}. The Einstein frequency diminishes rapidly with
increasing $\kappa$, similarly to the 3D case \cite{KRDW00,OH2000}.

\begin{figure}
\begin{center}
\epsfxsize=7cm
\epsfbox{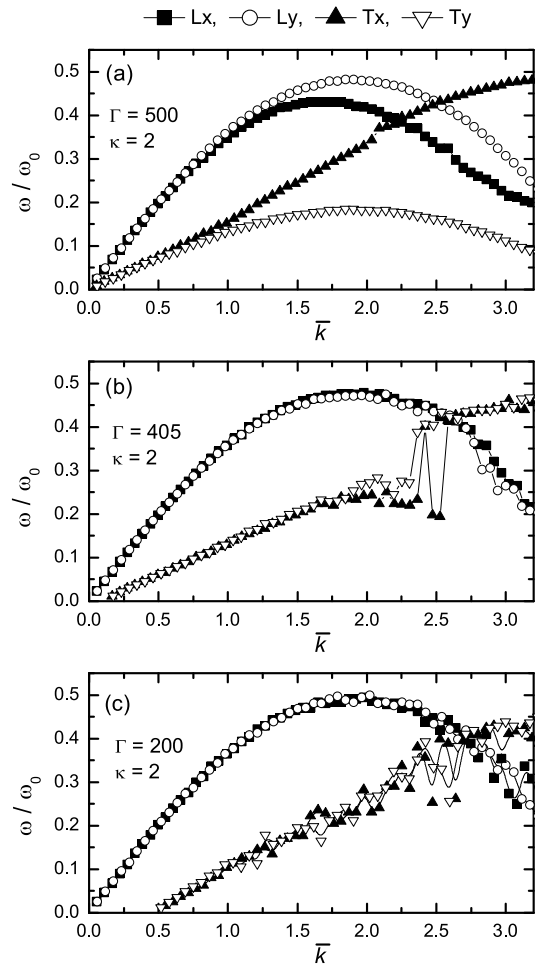}
\end{center}
\caption{\label{fig:2ddispmelt} 2D Yukawa systems: Comparison of MD (L
  and T) dispersions in the solid phase ($\Gamma=500$), just below the
  melting transition ($\Gamma=405$) and in the liquid phase
  ($\Gamma=200$) for $\kappa=2$. Shown are both $x$ and $y$
  polarizations, where $x$ is in the direction to the nearest neighbor
  in the hexagonal lattice.  Reproduced from
  Ref. \cite{IEEE07b}. Copyright (2007) by the Institute of Electrical
  and Electronics Engineers.}
\end{figure}

It is of interest to follow the evolution of the mode structure across
the liquid-solid phase boundary, as the isotropic liquid dispersion
transits into the anisotropic dispersion of the solid state. This is
illustrated for the $\kappa=2$ case in \fref{fig:2ddispmelt}.
$\Gamma=500$ represents a relatively high temperature solid, where
lattice defects may already show up, but the overall behavior (sharp
separation of the mode frequencies along the $x$ and $y$ directions;
compare e.g. the curves labeled Tx and Ty) reflects the conservation
of the triangular crystalline structure.  The $\Gamma=405$ case
corresponds to a temperature slightly higher than the melting
temperature (our results indicate that the transition occurs at
$\Gamma \cong 415$ for $\kappa=2$ \cite{IEEE07b}), where all long range
order in the system has already been extinguished, but locally most of
the particles sit in the somewhat distorted hexagonal environment. The
``oscillatory'' feature in the $T$ mode around $\bar{k}=2.5$ can be
taken as an indication for the transition from the ordered lattice to
the disordered liquid state through the formation of disoriented
domains of local hexagonal order. The orientation of these domains
becomes more uncorrelated with increasing temperature.  The
$\Gamma=200$ system is a typical strongly coupled liquid. Most
prominent features are the isotropy of the dispersion ($x$ and $y$
directions are equivalent), and the appearance of a finite wavenumber
cut-off for the $T$ mode.

\begin{figure}
\begin{center}
\epsfxsize=13cm
\epsfbox{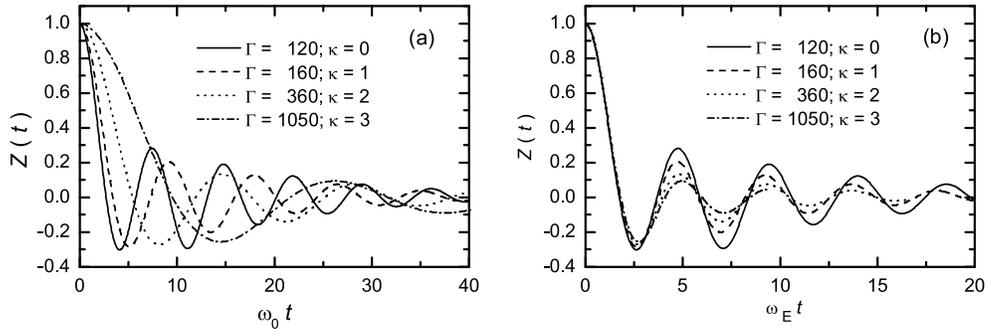}
\end{center}
\caption{\label{fig:y2dvac0} 2D Yukawa and Coulomb liquids: (a) velocity
autocorrelation functions for a series of $\kappa$ values. (b) The
same data as a function of $\omega_{\rm E} t$. The $(\Gamma,\kappa)$
pairs correspond to the same effective coupling $\Gamma^\star$.
Reproduced from Ref. \cite{H2005}.  Copyright (2005) by the American
Physical Society.}
\end{figure}

\begin{figure}
\begin{center}
\epsfxsize=13cm
\epsfbox{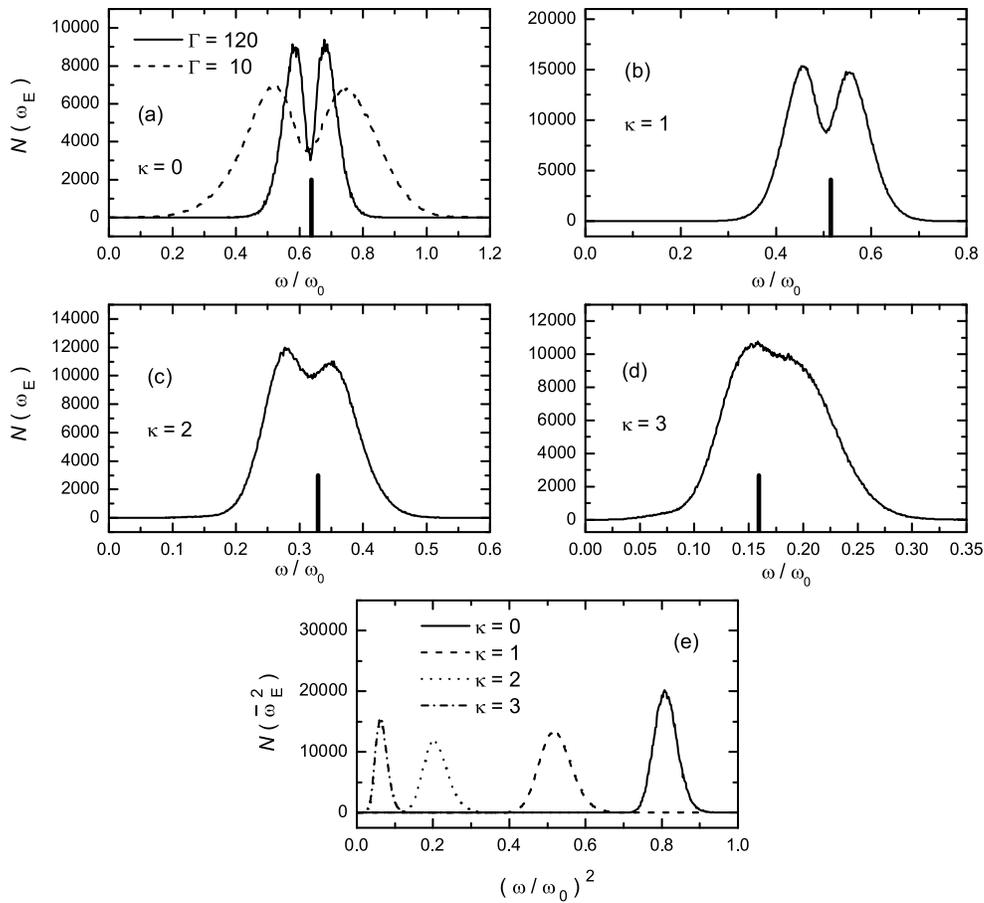}
\end{center}
\caption{\label{fig:2dMDeindisp}2D Yukawa and Coulomb liquids: 
Einstein frequency
distributions for $\Gamma^\star = 120$: (a) $\Gamma = 120$,
$\kappa=0$; (b) $\Gamma = 160$, $\kappa=1$; (c) $\Gamma = 360$,
$\kappa=2$; (d) $\Gamma = 1050$, $\kappa=3$; the vertical bars
indicate values obtained by the QLCA theory. (a) also shows the
frequency distribution obtained at a lower coupling value $\Gamma =
10$.  (e) Distribution of $\bar{\omega}_{\rm E}^2$ for the same
systems.}
\end{figure}

The behavior of the velocity autocorrelation function $Z(t)$ of the 2D
liquid is very similar to its 3D counterpart, at least for short
times. Representative $Z(t)$ functions obtained at different screening
parameter values are shown in \fref{fig:y2dvac0}(a). These functions,
when plotted against $\omega_{\rm E} t$ exhibit nearly universal
behavior [see \fref{fig:y2dvac0}(b)], indicating the relevance of the
Einstein frequency in determining the single particle properties
\cite{H2005}. While the in-depth analysis of the long-time behavior of
the velocity autocorrelation function is beyond the scope of this
paper, it is noted that in low-dimensional systems $Z(t)$ may exhibit
a slow power law decay, which makes it non-integrable
\cite{Alder70}. As a consequence the diffusion coefficient may not
exist for some 2D systems. The case of 2D Yukawa liquids has attracted
considerable attention during the last years
\cite{sdiff2007,Goree2008,Ott08,corr08}. These studies have found very
nearly $Z(t) \propto t^{-1}$ decay of the velocity autocorrelation
function and superdiffusion to exist for some conditions.

Similarly to the 3D case, numerical experiments were also performed
for the 2D case to determine the distribution of the microscopic
Einstein frequencies. A series of frequency histograms for
$\Gamma^\star=120$ at different values of $\kappa$ are shown in
figures~\ref{fig:2dMDeindisp}(a)-(d). We observe two peaks in the
distributions, which gradually get wider with increasing $\kappa$. The
QLCA results are again in very good agreement with the mean values of
the distributions. A wider frequency distribution appears when
$\Gamma$ is lowered [see \fref{fig:2dMDeindisp}(e)].  The
distributions of the sums of the 2 microscopic squared Einstein
frequencies $\bar{\omega}_{\rm E}^2$ -- as shown in
\fref{fig:2dMDeindisp}(e) -- are in contrast with the 3D
situation. Here we do not find qualitative difference between the
$\kappa=0$ Coulomb and the $\kappa \ne 0$ Yukawa cases for reasons
discussed in section \ref{sec:theory}.

\begin{figure}
\begin{center}
\epsfxsize=8cm
\epsfbox{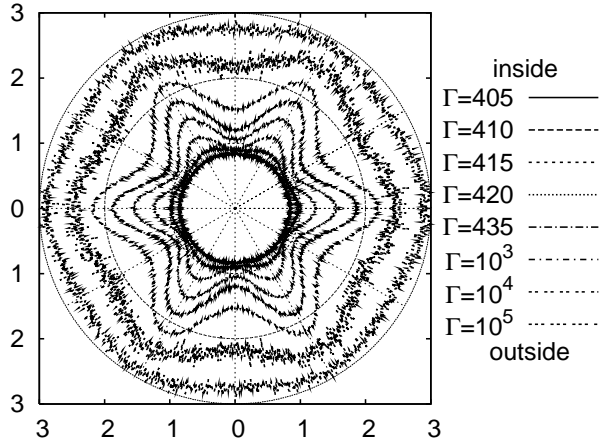}
\end{center}
\caption{\label{fig:snowflake}2D Yukawa systems: distribution of the
polarization angle for the higher frequency normal mode for different
values of the coupling parameter $\Gamma$ across the crystallization
boundary. $\kappa = 2$. Reproduced from Ref. \cite{IEEE07a}.
Copyright (2007) by the Institute of Electrical and Electronics
Engineers.}
\end{figure}

\begin{figure}
\begin{center}
\epsfxsize=13cm
\epsfbox{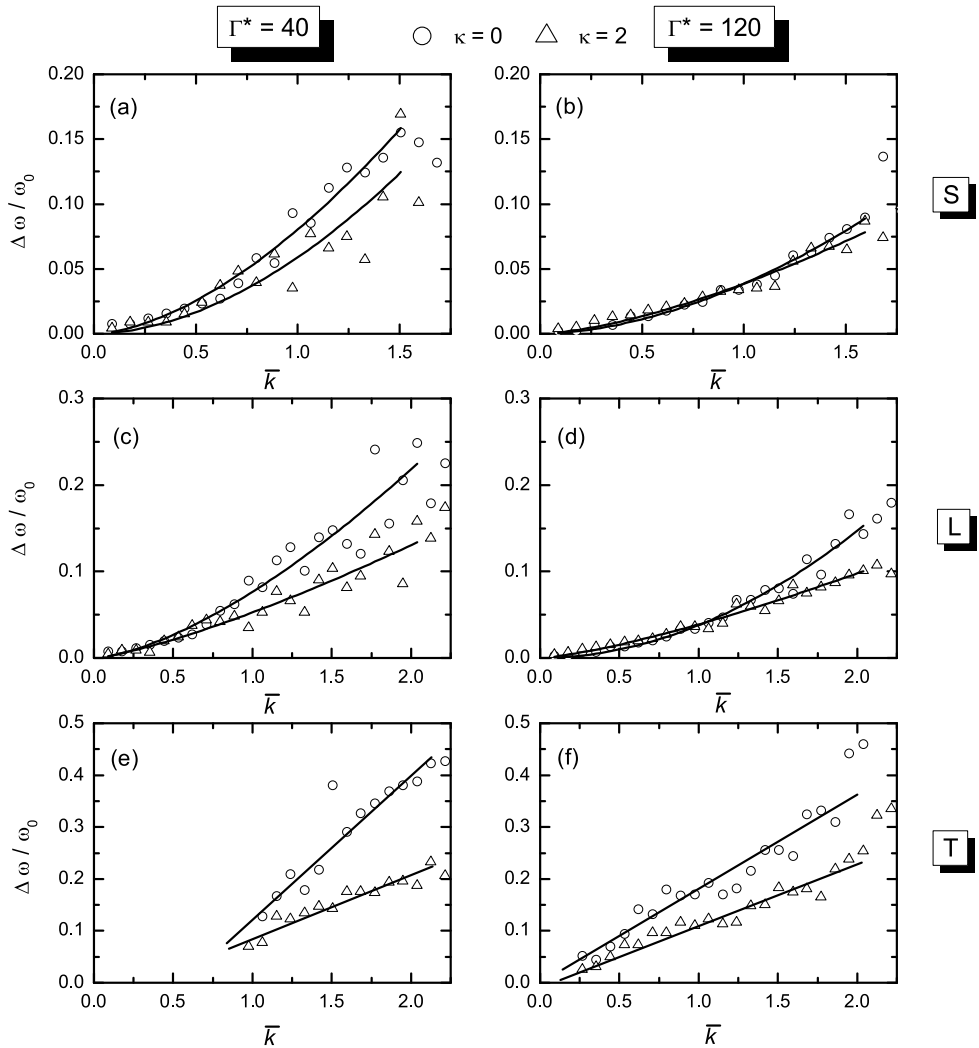}
\end{center}
\caption{\label{fig:2dspecwidth}2D Yukawa and Coulomb liquids: 
spectral line width
for $\kappa=0$ and $\kappa=2$ at $\Gamma^\star=40$ (a,c,e) and
$\Gamma^\star=120$ (b,d,f). Shown are FWHM (full-width at
half-maximum) values for the most prominent peaks in the
$S(k,\omega)$, $L(k,\omega)$ and $T(k,\omega)$ spectra.}
\end{figure}

\begin{figure}
\begin{center}
\epsfxsize=7cm
\epsfbox{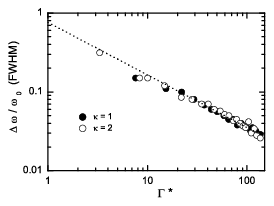}
\end{center}
\caption{\label{fig:gammadep} 2D Yukawa and Coulomb liquids: line
widths of the $S(k,\omega)$ spectra as a function of the reduced
coupling parameter $\Gamma^\star$, at a fixed wave number $\bar{k} =
1.0$.}
\end{figure}

The angular distribution of the polarization vector of the higher
frequency normal mode oscillation has also been analyzed, as an
indicator of the prevailing disorder
\cite{IEEE07a}. \Fref{fig:snowflake} shows the distribution of the
polarization angle for the higher frequency normal mode for different
values across the crystallization boundary (at $\Gamma \cong 415$).
As discussed in section \ref{sec:theory}, both the liquid (away from
the phase transition boundary) and the perfect lattice (extremely high
$\Gamma$ values) exhibit a full rotational symmetry, while in between
the sixfold symmetry of the lattice prevails.

The diagram showing the widths of the spectra of the dynamical
structure functions is displayed in \fref{fig:2dspecwidth}. Comparison
with data for the 3D Yukawa liquid reveals that the trends and
orders-of-magnitudes in the two cases are not substantially
different. The dependence of the width of the peaks in the
$S(k,\omega)$ spectra as a function of the reduced coupling parameter
follows the form $\Delta \omega / \omega_0 \cong 0.76
(\Gamma^\star)^{-2/3}$, as it is shown in \fref{fig:gammadep}. This
monotonically decreasing function of the coupling may change character
at lower coupling values, for which data are at present not
available. This is expected on the basis of the prediction by Hansen
{\it et al.} \cite{HMP75} (for the 3D case): for weak coupling the
width is expected to increase from its Vlasov value where
$S(k,\omega)$ should be extremely sharp, since higher coupling leads
to higher collision frequency and thus to stronger damping. Once,
however, localization sets on, further increase in the coupling is
expected to create better localization and thus a reduction in the
collision frequency and in the width. Thus, generation of data for
lower $\Gamma$ values would be desirable, to see whether a turnaround
point really exists.

\subsection{Quasi-two-dimensional Yukawa liquids confined by a parabolic potential} 
\label{sec:qtwoD}

The model adopted for the 2D Yukawa system, which assumes that the
particles are constrained to move entirely within an ideal plane can
be extended to describe more accurately the situation found in
physical systems, by allowing small amplitude displacements of the
particles perpendicular to the plane. In this extended model one
applies a parabolic potential along the direction perpendicular to the
plane, which then results in a \textit{quasi-two-dimensional}
confinement. Such confinement gives rise to a particle layer with
finite width, or -- at weaker confinement -- to a sequence of
multilayer structures, when the confinement or interaction potential is
varied. The structural phase transitions (a change in the number of
layers and in the accompanying crystal structures), relevant to
particle traps, have theoretically been studied by Dubin \cite{Dubin},
while Totsuji {\it et al.} \cite{Totsuji1}, Bystrenko \cite{Bystrenko}
as well as Qiao and Hyde \cite{QH05} investigated the formation of
layers in Coulomb and Yukawa systems in confined quasi--2D
configurations.

The number of layers formed in the liquid phase depends on the
strength of the confinement. In contrast to the idealized 2D systems
the layers have a finite width. Here we deal with the domain of
parameters when a {\it single layer} is formed. At higher number of
layers the mode structure is expected to be more complex
\cite{DKHGK_prl,CPP,GK03}, but the study of these modes is not within
the scope of the present analysis. In a single-layer configuration
the third degree of freedom of the particles, the displacement
perpendicular to the plane, gives rise to an additional collective
excitation, the ``out-of-plane'' $\cal{P}$ mode, besides the
``in-plane'' $\cal{L}$ and $\cal{T}$ modes found in (ideal) 2D layers. 
The out-of-plane mode in the
crystallized state has been studied through simulations by Qiao and
Hyde \cite{QH03}. Results pertaining the strongly coupled liquid phase
were analyzed in \cite{DHK_out}, and will be summarized below.  It
should also be noted that a somewhat similar physical situation arises
when a 1D chain of particles is confined in the transverse direction
by a parabolic potential: indeed, there is a similarity between the
modes that represent excursions along the direction of the confining
force in the 1D and 2D systems.

In the quasi-2D liquid system the particles can freely move in the
$(x,y)$ plane while a confinement potential $V_c(z) \propto z^2$
acts upon them when they are displaced from the $z$ = 0 plane. The
confinement force is linear with respect to the ``vertical''
displacement,
\begin{equation}\label{eq:force} 
F_z= - f_0 \frac{Q^2}{4 \pi \varepsilon_0 a^3} z,
\end{equation}
where the strength $f_0$ (besides $\Gamma$ and $\kappa$) is the third
characteristic parameter of the system. At $f_0 =1$ the confinement
force at a vertical displacement $z=a$ is equal to the magnitude of
the force between two particles separated by $a$ [defined by
(\ref{eq:aws2d})], interacting via Coulomb potential.  Information
about the (thermally excited) collective modes and their dispersion is
obtained from the analysis of the correlation spectra of the
longitudinal and (in-plane as well as out-of-plane) transverse current
fluctuations. For the ``in-plane'' $\cal{L}$ and $\cal{T}$ modes we
use eq. (\ref{eq:dyn}), while for the out-of-plane mode the
corresponding microscopic current $\pi(k,t)$ (which characterizes the
$\cal{P}$ mode) is obtained as:
\begin{eqnarray}
\pi(k,t)= k \sum_j v_{j z}(t) \exp \bigl[ i k x_j(t) \bigr].
\label{eq:dyn_pi}
\end{eqnarray}

\begin{figure}
\begin{center}
\epsfxsize=13cm
\epsfbox{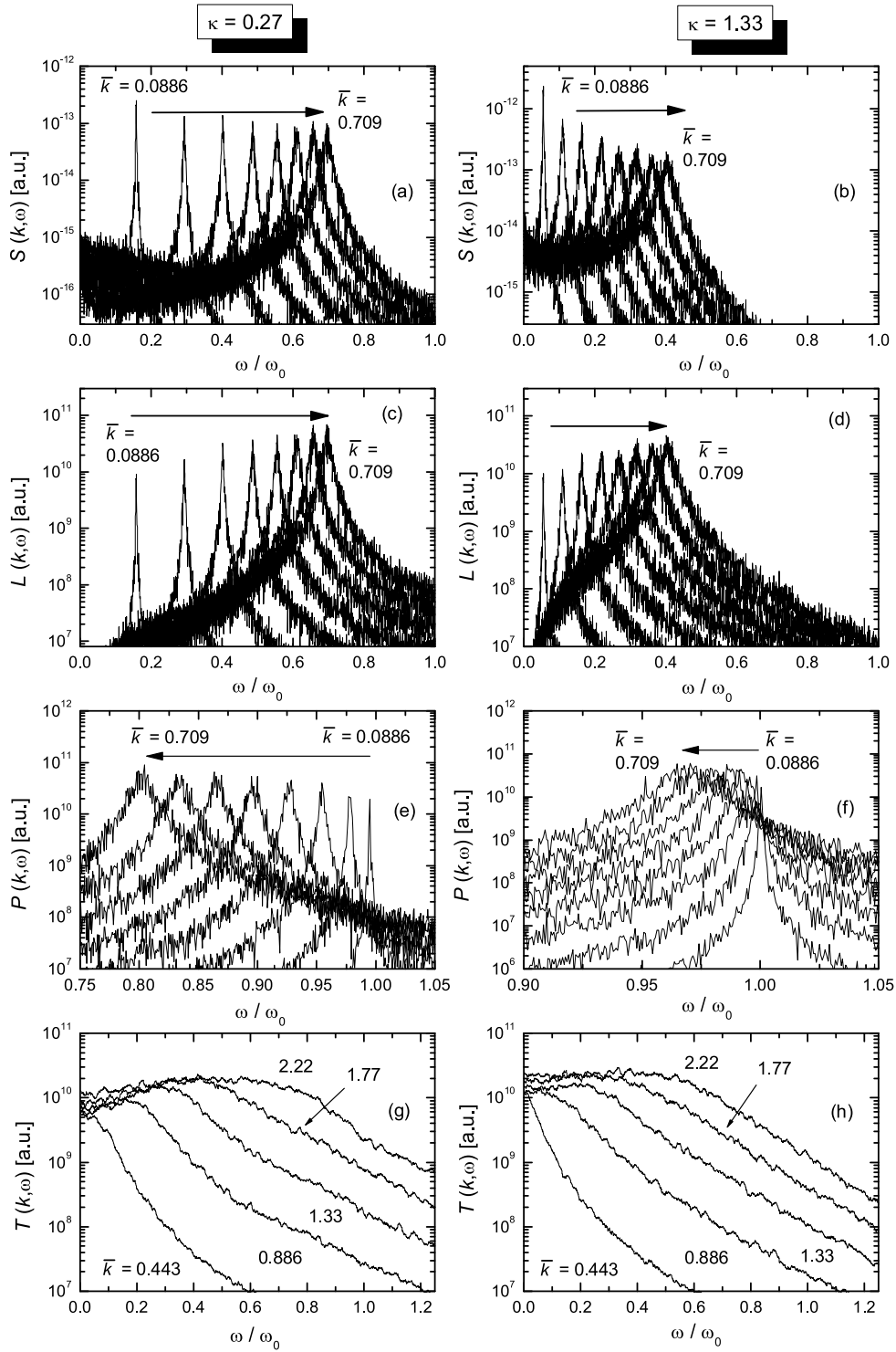}
\end{center}
\caption{\label{fig:outspectra}Quasi-2D Yukawa liquids: dynamical
  structure function [$S(k,\omega)$], longitudinal [$L(k,\omega)$]
  and out-of-plane as well as in-plane transverse [$P(k,\omega)$ and
  $T(k,\omega)$] current fluctuation spectra for $\kappa$ = 0.27
  (a,c,e,g) and $\kappa$ = 1.33 (b,d,f,h). The $S$, $L$, and $P$
  spectra are plotted for the multiples of the smallest accessible
  wave number $\bar{k}_{\rm min}$ = 0.0886, while the $T$ spectra are
  shown for higher wave numbers indicated by the labels in (g) and
  (h). The arrows in (a)-(f) indicate increasing wave numbers. Confining force:
  $f_0=2$. Reproduced from Ref. \cite{DHK_out}. Copyright (2004) by
  the American Physical Society.}
\end{figure}

Representative current fluctuation spectra for the three ($\cal{L}$,
$\cal{P}$, and $\cal{T}$) modes are displayed in
\fref{fig:outspectra}, for $\Gamma$ = 100, $f_0$ = 2.0 and two
different values of the screening parameter $\kappa$ = 0.27 and
$\kappa$ = 1.33. The frequency is normalized according to
(\ref{eq:omegap2d}). We observe sharp peaks in the $L(k,\omega)$
spectra, similarly to the case of (ideal) 2D Coulomb and Yukawa
liquids \cite{MG03,KHDR_prl}, characteristic of long-lifetime
collective excitations \cite{Vladimirov97}. Peaks in the $\cal{T}$
mode spectra [see \fref{fig:outspectra}(g,h)] show up only above a
certain (cutoff) wave number, similarly to the case of 2D and 3D
Yukawa systems, as discussed in the previous sections.

The $\cal{P}$ mode possesses a finite frequency at $k=0$, which is,
in general, characteristic of an {\it optical} mode. The first
identification of this pseudo-optical behavior in a confined 2D system
is due to \cite{Vladimirov97}. At small wave numbers the
peaks of the spectra shift to lower $\omega$ as $\bar{k}$ is
increased. The width of the peaks of the $P(k,\omega)$ spectra become
gradually broader when $\kappa$ is increased, as it can be seen in
\fref{fig:outspectra}(e) and (f). It is noted that, on the other hand,
the peaks become narrower as the strength of the confining potential,
$f_0$, is increased, which is an indication of an increasing lifetime
of this collective excitation.

\begin{figure}
\begin{center}
\epsfxsize=13cm
\epsfbox{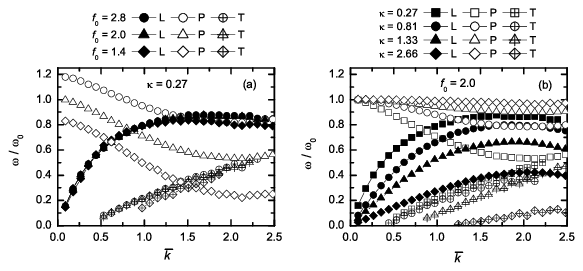}
\end{center}
\caption{\label{fig:outdisp}Quasi-2D Yukawa liquids: dispersion
  relations for (a) $\kappa$ = 0.27 and different values of the
  amplitude $f_0$ of the confining potential, and (b) for fixed $f_0$
  = 2 and different values of the screening parameter.  Reproduced
  from Ref. \cite{DHK_out}. Copyright (2004) by the American Physical
  Society.}
\end{figure}

\begin{figure}
\begin{center}
\epsfxsize=6cm
\epsfbox{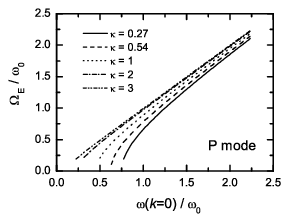}
\end{center}
\caption{\label{fig:out-einstein}Quasi-2D Yukawa liquids: the
relation between the frequency $\omega(k=0)$ and the Einstein
frequency of the $\cal{P}$ mode for different values of the screening
parameter $\kappa$.  Reproduced from Ref. \cite{DHK_out}. Copyright
(2004) by the American Physical Society.}
\end{figure}

The dispersion relations derived from the spectra are displayed in
\fref{fig:outdisp} for different values of $f_0$ and $\kappa$ for
$\Gamma$ = 100. At constant $\kappa$, as shown in
\fref{fig:outdisp}(a), the frequency of the out-of-plane mode changes
significantly as the strength of the confinement force, $f_0$, is
varied. The $\cal{L}$ and $\cal{T}$ modes are only slightly affected
by the value of $f_0$. The frequency of these modes is somewhat
smaller at $f_0$ = 1.4, which is near the lower bound of $f_0$ for the
formation of a single layer \cite{DHK_out}. It is noted that at lower
$f_0$ values, when two layers are formed, two longitudinal and two
in-plane transverse modes appear, similarly to those identified in the
classical (ideal) bilayer system \cite{DKHGK_prl}.  Additionally, two
out-of-plane transverse modes also emerge in the two-layered system,
which are also believed to be in-phase and out-of-phase modes (when
particles in the two layers oscillate in phase or with a phase
difference of $180^{\circ}$ in the two layers). The $\cal{L}$ mode
exhibits a quasi acoustic behavior, with a linear portion of the
dispersion curve around $k$ = 0, which widens with increasing
$\kappa$, as it can be seen in \fref{fig:outdisp}(b). The $\cal{T}$
mode shows an acoustic, $\omega \sim k$ dispersion at small $k$, with
a cutoff at a finite wave number.

For the $\cal{P}$ mode ${\rm d}\omega / {\rm d}k < 0$ in the $\bar{k}
\lesssim 2.1$ domain. At higher wave numbers, the frequency of the
mode slightly increases with $k$. This observation on the liquid
system agrees well with that on the crystallized system \cite{QH03},
where the same behavior was found, except that in the latter system
the critical wave number (at which the group velocity ${\rm d}\omega
/ {\rm d}k$ changes from negative to positive) also depends on the
direction of the propagation.

At $k = 0$ the whole layer oscillates in unison in the potential well
with a frequency:
\begin{equation}
\frac{\omega(k=0)}{\omega_0} = \sqrt{f_0/2}. \label{eq:omega1}
\end{equation}
A smaller confinement force results in a smaller
$\omega(k=0)$ and $\omega(k \rightarrow \infty)$. At a constant $f_0$
the value of $\omega(k=0)$ does not change when $\kappa$ is varied,
but -- as shown in \fref{fig:outdisp}(b) -- $\omega(k>0)$ increases
with decreasing $\kappa$. This is explained by the decreased
interparticle force (at an average particle separation) at higher
$\kappa$.

The frequency of the out-of-plane mode at the $k \rightarrow \infty$
limit (i.e. the Einstein frequency \cite{DHK_aps,Bakshi}) can be
calculated by considering the forces acting upon a single particle
displaced in the $z$-direction, while all other particles are in rest
in the $z$ = 0 plane. The force is the sum of the confining force and
the force due to repulsion by the other particles,
\begin{equation}
\label{q1} F(z)= -f_0 \frac{Q^2}{4 \pi \varepsilon_0 a^3} z + F_{\rm r}(z). 
\end{equation} 
The $F_{\rm r}(z)$ contribution can be
calculated as $F_{\rm r}(z) = - \partial V_{\rm r} / \partial z$,
where $V_{\rm r}(z)$ is the potential distribution due to a charge
distribution $\rho(x,y)$ in the $z$ = 0 plane. To obtain $\rho(x,y)$
one may either use the radial (2D) pair correlation function (PCF) or
consider the particles occupying hexagonal lattice sites in the $z=0$
plane. The $F_{\rm r}(z)$ force is found to be a nearly linear
function of the displacement $z$, in the $|z| < 0.3 a$ domain, where
the particle displacement is expected to fall. The resulting
(Einstein) frequency (when the particles in the $z=0$ plane are
situated at lattice sites) is \cite{DHK_out}:
\begin{equation}\label{eq:highk}
\frac {\Omega_{\rm E}} {\omega_0} = \frac {\omega(k \rightarrow \infty)}
{\omega_0} \cong \sqrt{\frac{f_0 - 1.63 \exp(-1.37 \kappa)}{2}}.
\end{equation}
In the case of using the disordered configuration in the $z$ = 0
plane instead of lattice sites (through PCFs obtained in the liquid
state simulations), a frequency very close to that given by
(\ref{eq:highk}) is obtained. At low values of $\kappa$ the Einstein
frequency $\omega_{\rm E}$ is significantly lower than
$\omega(k=0)$, as illustrated in \fref{fig:out-einstein}. In the
high $\kappa$ limit the two frequencies are equal, as the screening
becomes very strong and the particles interact very weakly. In this
case the frequency of the $\cal{P}$ mode becomes nearly independent
of $\bar{k}$.

\section{Experimental results}\label{sec:experiment}

Experimental results on wave propagation and collective excitations in
Yukawa systems have been accumulating in complex (dusty) plasma
experiments since the mid 1990--s. Experiments have been carried out
both on spontaneously generated and on externally excited waves.  An
early laboratory observation of longitudinal modes was reported by
Barkan {\it et al.} \cite{Barkan95} in 1995, followed by a more
detailed study in 1997 \cite{Thompson97}: these authors observed
spontaneously generated waves in dust that filled a volume with a
cylindrical geometry. These waves grew as the result of the
dust-acoustic instability, which was driven by an ion flow, and the
experimenters were able to measure the wavelength and propagation
speed. An early effort to excite waves by manipulation using an
electrically-biased wire was reported by Pieper and Goree
\cite{Pieper96}, who also introduced a method of data analysis that
yields the real and imaginary parts of the wave number, for the
applied frequency. Repeating the measurements at various frequencies
yielded a dispersion relation. In some of these early experiments the
ambient pressure was kept high in order to avoid instabilities. As a
result, as pointed out by Rosenberg and Kalman \cite{RosKal97}, the
waves were strongly damped, primarily by grain-neutral
collisions. Thus, even though the experiments were conducted under
strongly coupled conditions in the liquid state, the strong
collisional damping washed away the difference between weakly coupled
and strongly coupled dispersions (see \cite{Kaw2,RosKal97} for a more
detailed discussion). Two recent experiments have further corroborated
this picture. Bandyopadhyay {\it et al.} \cite{Bandy07} investigated
the acoustic dispersion of the longitudinal mode in the strong
coupling regime over a wide range of the neutral pressure values and
found $\partial\omega / \partial k < 0$ behavior of the dispersion
curve that in the low collisional domain could be attributed to
correlational effects. On the other hand, the experiment reported by
Annibaldi {\it et al.}  \cite{Annibaldi07}, in the high collisional
regime confirmed that in this domain the strong coupling effects were
washed away completely.

The first experiments where strong coupling effects were clearly
displayed were done on a 1D complex plasma in the crystalline state,
realized as a chain of grains held together in the transverse
direction by a confining potential. Longitudinal waves (along the
direction of the chain) excited by the radiation pressure of a laser
beam \cite{Homann97,Peters96} in a parallel plate
radio frequency discharge were observed: the analysis led to the
conclusion that the weakly coupled theory of the longitudinal waves
(referred to as ``dust acoustic waves'' \cite{Rao90}) was
inadequate, while the description in terms of harmonic phonons of a
system with short range interaction (referred to as ``dust lattice
waves'' \cite{Melandso96}) provided a more satisfactory agreement
with experiments.

Generation of 3D complex plasmas in the laboratory under strong
coupling conditions and at sufficiently low pressure, so that strong
coupling effects become manifest turned out to be difficult. A good
summary of the state of affairs as of 2000 is given by
\cite{Kaw2}. This paper and a later work \cite{Kaw} also discuss the
spontaneous excitation of shear wave-like structures in a strongly
coupled liquid at a low pressure. The plasma originally was in a 3D
configuration, but assumed a layer structure in the course of the
experiment, with particle excursions in the direction perpendicular to
the layers. Thus it seems difficult to judge whether the observed
waves were indeed shear waves or some more intricate excitation in the
coupled layer system. 

The presence of the ion beam traversing the dust plasma generated in
low-pressure gas discharges can create issues that can not be
approached within the Yukawa model. Because of the anisotropy
introduced by the ion beam, the Yukawa interaction will be modified in
the vertical direction (along the beam), but probably not too much in
the horizontal plane \cite{Lemons,Lapenta}. This scenario was
supported by experiment
\cite{MorfillPOP,Steinberg01,Hebner03,Hebner04}. A further major
problem due to the ion beam in the 3D geometry was identified by Joyce
{\it et al.} \cite{Joyce02}, that in the low pressure domain, where
collective modes could be observable, ion--dust instability may lead
to melting. To avoid these problems, most of the subsequent laboratory
experiments were to favor 2D geometries over three-dimensional ones:
in a 2D system these problems should be absent. Since 1998 substantial
progress in the understanding of the excitation and propagation of
waves in 2D Yukawa systems has ensued. In the strong coupling regime
the constituent grains are, in principle, either in the crystalline
solid or in the liquid state. In fact, in addition to the formation of
large scale ordered lattice structures a more common configuration
is an aggregate of micro-crystals whose prevailing disorder is
expected to make the behavior of the aggregate quite similar to that
of the liquid state.

Longitudinal waves in a 2D dust plasma crystal were first observed
experimentally in a parallel plate radio frequency discharge by Homann
{\it et al.}  \cite{Homann98}. The observation of transverse
(in-plane) shear waves, the hallmark of strong coupling, excited by a
chopped laser beam was reported by Nunomura {\it et al.}
\cite{NSG00}. Their measurements of the dispersion relation revealed
an acoustic, i.e., non-dispersive, character over the entire range of
wave numbers measured, ($0.3<\bar{k}<1.2$), at $\kappa \approx 0.74$,
with transverse sound speed and Einstein frequency values in agreement
with theory \cite{Sullivan,Peeters87}.

A series of beautiful experiments on the generation of Mach cones in
the wake of an object moving through a 2D dusty plasma crystal was
also crucial albeit in an indirect way, in determining the strong
coupling characteristics of these systems. It was unambiguously shown
\cite{Samsonov99,Samsonov00,Melzer00} that Mach cones appear when the
velocity of the moving object (particle or laser spot) exceeds the
longitudinal sound speed $s_{\rm L}$ in the medium. Subsequent
observations \cite{Nosenko02} with object velocities below this limit,
but above the transverse shear sound speed $s_{\rm T}$ (the ratio of
the two speeds had the value $s_{\rm L}/s_{\rm T}=4.48$ in the
experiment) also demonstrated the excitation of a small angle Mach
cone sustained by the transverse mode.

\begin{figure}[h!]
\begin{center}
\epsfxsize=6cm
\epsfbox{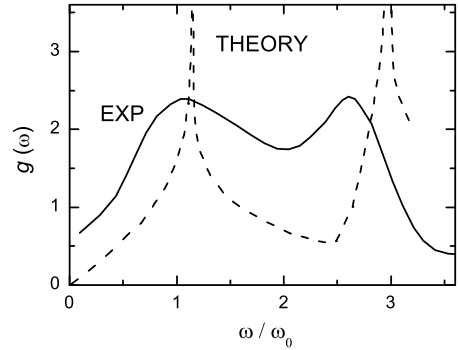}
\end{center}
\caption{\label{fig:gomega}2D system: density of states as obtained from
the experiment of Nunomura {\it et al.} \cite{Nunomura02} and by theory.}
\end{figure}

\begin{figure}[h!]
\begin{center}
\epsfxsize=13cm
\epsfbox{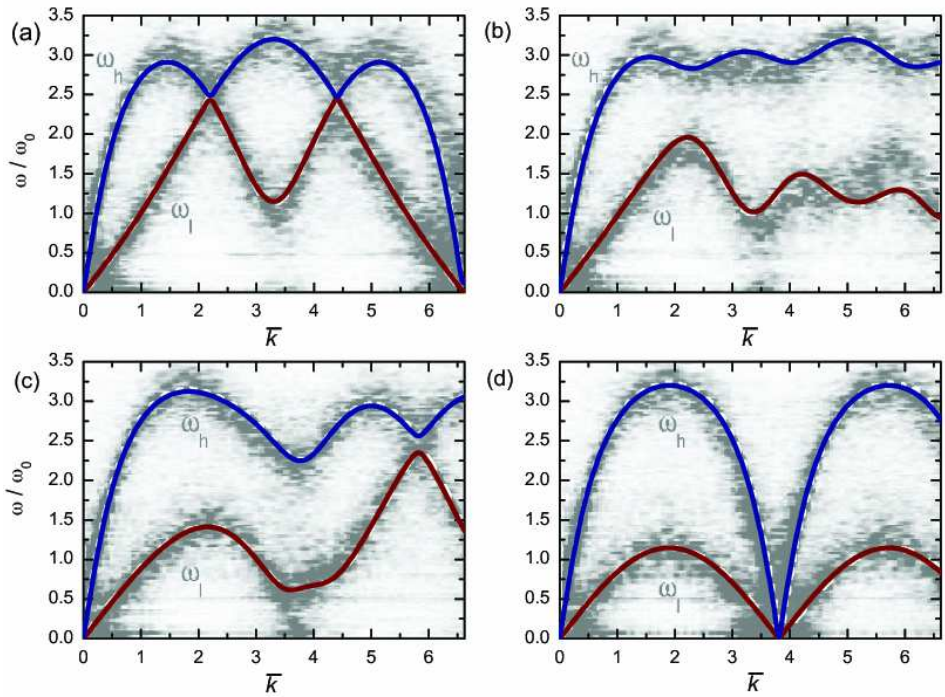}
\end{center}
\caption{\label{fig:expdisp}(color online) 2D system: wave dispersion
in directions (a) 0, (b) 10, (c) 20, and (d) 30 degrees (measured from
the nearest neighbor direction) as obtained in the experiment of
Zhdanov {\it et al.} \cite{Zhdanov03} and calculated from lattice
summation (heavy lines). Experimental data reproduced from
Ref. \cite{Zhdanov03} with kind permission of the authors. Copyright
(2003) by the American Physical Society.}
\end{figure}

More recent experiments done both on the 2D solid and liquid phases
have been able to determine plasma parameters with sufficient accuracy
and to perform measurements of great number of observables, so that
detailed quantitative comparisons with the theoretical predictions
have become possible. Experiments by Nunomura {\it et al.}
\cite{Nuno02} introduced laser manipulation, which avoided technical
problems caused earlier by the electrical wires, and the dispersion
relations were measured with greater accuracy. Subsequently Nunomura
{\it et al.} \cite{Nunomura02} detected the spectra of self-generated
longitudinal and transverse excitations along the two principal axes
of a triangular lattice. The energy was concentrated along a well
defined $\omega(k)$ curve, representing the measured dispersion
relation. The data covered the $0<\bar{k}<3.3$ domain with $\kappa
\approx 0.74$; our comparison with the theoretical dispersion curves
calculated by Peeters and Wu \cite{Peeters87} and by Sullivan {\it et
al.} \cite{SullivanUP} shows excellent agreement with these data. A
partial frequency spectrum (i.e. density of states), based on the
kinetic energy contents of the 4 selected modes was also generated:
while comparison with the calculated spectrum (see section
\ref{sec:theory}) is possible, agreement beyond what is visible in
\fref{fig:gomega} is not expected, since the theoretical spectrum
includes all propagating modes \cite{Schwabe07}. In a subsequent work
\cite{Zhdanov03} the spectrum of waves was measured also for a number
of directions in between the principal axes and over a much broader
domain of wave number values: $0 < \bar{k} < 6.6$. In addition to the
frequencies, the polarization angles of the modes were also determined
(the mode polarizations can be described as ``longitudinal'' and
``transverse'' for propagation along the principal directions
only). All these data show excellent agreement with theory
\cite{Peeters87,SullivanUP}, see \fref{fig:expdisp}.

Following a different line of approach Melzer \cite{Melzer03} studied the
normal modes of small 2D clusters of grains; what is of interest in
the present context is the transition from the mode spectrum of a
finite number of particles into that of an ``infinite'' system. With a
somewhat arbitrary assignment of labels for the normal modes, it was
found that the average frequency as a function of $k$ provides a fair
resemblance to the $\omega(k)$ dispersion of the infinite lattice,
already for a cluster as small as consisting of 34 particles. The
scatter of frequencies around the $\omega(k)$ curve is of course
substantially higher than in the experiments quoted above. (For
somewhat related results see \cite{DKG_caging}). The theoretical
understanding of the distribution of dynamical frequencies (as a
function of temperature and particle number) presents a major
theoretical challenge, with a very limited body of antecedents
available in the literature \cite{Elliot,Czahor}. The possibility of
generating experimentally observable scenarios from which information
on the frequency distribution can be extracted should provide stimulus
for new theoretical efforts. 

As to the liquid state, observational data available at the present
time are quite recent and still rather limited.  Nunomura {\it et al.}
\cite{Nunomura05} studied the change of the thermally excited mode
structure as the crystal lattice was melted and the system transited
to the strongly coupled liquid state. The melting was achieved by
directed laser heating. The theoretically predicted trends, such as
the development of a cut-off wave number for the shear mode and the
shift of the longitudinal mode frequency towards higher values are in
fair agreement with MD data. On the other hand, the widths of the
spectra seem to be higher than expected. It is difficult to relate the
results of the MD studies of the break-up and isotropization of the
lattice modes \cite{IEEE07b} to this experiment, since the $\Gamma$
values where the observations of the liquid state were done are quite
far from the phase transition point.
 
A careful study of the transverse modes in the strongly coupled liquid
state, in the vicinity of the melting point is due to Piel {\it et al.}
\cite{Piel06}. These authors analyzed the propagation of externally
excited shear waves, through a sophisticated data analysis technique
that made it possible to collect information from a high noise
environment. The experimental situation corresponded to $\kappa 
\approx 0.4$, which allowed the comparison with the QLCA data for
$\kappa=0$ and $\kappa \approx 1$ as lower and upper bounds. The
authors found that within the wave number domain investigated
($0<\bar{k}<2.5$) the overall agreement between experiment and the QLCA
model is quite satisfying. They note that even in the solid state the
waves assume characteristics resembling those in the liquid state
(angularly averaged dispersion), because the plasma crystal consists
of domains of different orientations.  For this reason there does not
seem to be too much change in the dispersion, as one passes from the
solid phase to the liquid phase. On the other hand, the damping is
substantially higher on the liquid side and becomes stronger for low $k$
values.

\begin{figure}[h!]
\begin{center}
\epsfxsize=7cm
\epsfbox{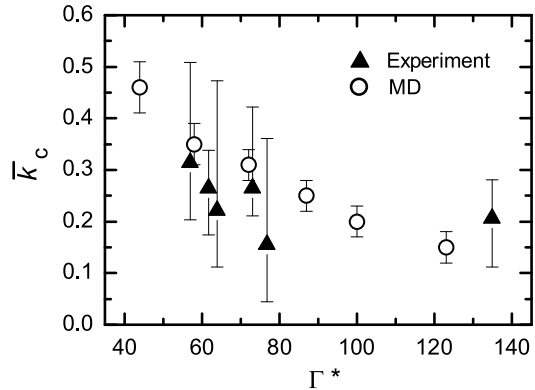}
\end{center}
\caption{\label{fig:cutoff2d}
2D Yukawa liquids: Transverse mode cutoff wavenumber
$\bar{k}_{\rm c}$. Experimental values are taken from \cite{Nosenko06}
and are compared with 2D Yukawa molecular dynamics results.}
\end{figure}

Since the homogeneous liquid cannot sustain shear, the shear mode must
vanish below some finite $k_{\rm c}$ value. The value of this cut-off wave
number was recently studied by Nosenko {\it et al.} \cite{Nosenko06}
in a low pressure experiment.  At the $\kappa \approx 0.43$ of the
experiment $\bar{k}_{\rm c}$ ranges between 0.16 and 0.31, as the coupling
strength $\Gamma$ is varied from the melting value $\Gamma=155$ down
to $\Gamma=60$. These values compare favorably with the values
obtained by the MD simulations reported in \cite{KHDR_prl} (see
fig.~\ref{fig:cutoff2d}). This can be taken as an indication that the
cut-off is attributable to the intrinsic dynamics of the grains. Thus
one can conclude that at low pressures the contribution of the
grain-neutral collisions to the generation of the cut-off is quite
negligible.

There are experiments on 1D chains of grains held together in the
transverse direction by a confining potential which reveal (in
addition to the observation of longitudinal waves on such systems
quoted above \cite{Homann97,Peters96}) excitations in the direction
perpendicular to the axis of the chain. These modes bear physical
features similar to the shear-waves in 2D systems. In particular, they
exhibit the benchmark pseudo-optic behavior and the ensuing negative
dispersion predicted by theory for the
latter \cite{QH05,QH03,DHK_out}. Experimentally, the mode
dispersion was determined by analyzing spontaneously excited waves by
Misawa {\it et al.} \cite{1da} and by creating the transverse waves
through the manipulation of a single particle by Liu {\it et al.}
\cite{1db,1dc} confirming these predicted features. In the 2D
geometry, self-excited out-of-plane oscillations of particles were
identified first by Nunomura \cite{Nunomura99}. Samsonov {\it et al.}
\cite{SZM05} investigated the propagation of wave packets in the
vertical (i.e. along the confinement) direction (see Section
\ref{sec:qtwoD} for more details) and confirmed the predicted
dispersion characteristics \cite{QH05,QH03,DHK_out} of the
out-of-plane ($\cal{P}$, pseudo-optic) mode.

\section{Summary} \label{sec:sum}

The objective of this review has been to summarize the huge body of
information that has been gathered since 1990-s through theoretical
analysis, computer simulations and laboratory experiments on the
collective excitations of dusty (complex) plasmas and from this to
determine  the collective behavior of two- and
three-dimensional strongly coupled Yukawa systems.  The Yukawa model
allows the mathematical analysis of an idealized system that
represents a variety of actual many-particle physical systems (dusty
plasmas, charged colloids, mesoscopic particles, etc.), which are
characterized by (i) a significant ratio of the potential energy
(originating from the high charge value of the particles) to the
kinetic energy in the system, as expressed through the plasma coupling
parameter $\Gamma$, and (ii) by a particle--particle interaction that is
strongly affected by a polarizable background coexisting with the main
plasma.

Two techniques have been used for the mathematical modeling: molecular
dynamics, as a computational simulation method and the Quasi-Localized
Charge Approximation as a theoretical scheme. The results generated by
the two independent approaches have been found to be in excellent
agreement with each other, and have been convincingly supported by the
findings of laboratory experiments. Thus all these assembled data
converge into a coherent and fairly complete physical picture, which
has been presented in this Review. Nevertheless, there are quite a few
areas that the reader may have expected to see in this paper, but
which have been excluded from consideration. Thus some qualifying
comments along this line are in order.
\begin{itemize}
\item{ 
  We have considered infinitely large, unbounded 2- and
  3-dimensional systems only: thus effects relating to 1D geometry,
  boundary conditions, inhomogeneities have been excluded. Some of
  these (Yukawa balls and disks e.g. \cite{Cball04,Sheridan}) have
  been attracting much attention lately.}
\item{ 
  A special configuration, familiar in semiconductor physics, is
  the bilayer geometry (consisting of two parallel 2D planes,
  separated from each other by a small distance $d$). While  
  semiconductor devices are governed by Coulomb interaction, a similar
  configuration is of interest with systems where Yukawa interaction 
  prevails \cite{LowenPRL}. The likelihood
  of the realization of such a geometry in laboratory dusty plasma
  experiments is not promising using identical grains, but may be more
  feasible in a microgravity environment. However, combining two 
  species of differently sized microparticles in a conventional 
  laboratory sheath geometry setup leads to the automatic formation 
  of a bilayer configuration, due to the different $Z/m$ ratios,
  as recently pointed out by Matthews {\it et al.} \cite{Matthews2006}
  and demonstrated in a subsequent experiment by 
  Smith {\it et al.} \cite{Smith2008}. Bilayer systems posses rich 
  variety of structural phases \cite{LowenPRL,Goldoni1996,Messina08}
  and a rather complex collective mode structure whose
  details exhibit a remarkable sensitivity to the layer separation
  \cite{DKHGK_prl,Matthews2006,qlca_gap}. Further experimental
  investigation of this behavior would be of great interest.}
\item{ 
  Only single component systems (the Yukawa equivalent of the
  OCP, one component plasma) have been treated; the crucially
  important extension to two- or multicomponent cases (different
  masses, different charges) is not here. Creation of such systems in
  the current laboratory set-ups is hampered by technical reasons (but 
  again may become feasible in a microgravity environment) and
  serious theoretical studies are lacking.}
\item{ 
  While we have presented simulation data of the dynamical
  fluctuation spectra in great detail, the evaluation and theoretical
  analysis of most of these data is still to be carried out. We have
  focused on the positions of the peaks in the spectra: the most
  important question amongst those whose analysis is incomplete is
  that of the widths of the peaks, which, in turn, are characteristic
  of the damping of the excitations. What is missing primarily is a
  solid theoretical foundation through which the different mechanisms
  that lead to the damping of the collective modes could be reconciled
  and against which the simulation data could be tested.  The QLCA
  analysis points at the main physical effects where the source of the
  damping should be sought, but no reliable analytic tool has emerged
  that would predict how the damping depends on the coupling strength
  and on the wavelength of the mode.}
 \item{ 
  The theoretical tool (the QLCA) described in this Review is
  geared to strong coupling and it provides no linkage from the
  localization dominated strongly correlated behavior to the fluid-like
  weakly correlated behavior of the collective modes. Only more
  simulation work and a different theoretical approach would bridge
  this gap.}
\item{ 
  There are both some experimental \cite{Zhdanov03} and
  simulation (see \fref{fig:spectra1} and \cite{H2005}) results
  available on the effect of the disorder on the collective mode
  spectrum. Both these and theoretical considerations suggest that one
  should think in terms of frequency distributions, rather than in
  terms of well-defined collective mode dispersions.}
\item{ 
  We have not discussed effects and phenomena relating to
  external or internally generated magnetic fields. These issues may
  become the topics of investigation for the next generation of
  complex plasma experiments. An externally imposed magnetic field
  could affect the polarizable medium (electrons and ions) and thus
  alter the effective interaction potential; at sufficient strength it
  may even change the orbits of the mesoscopic plasma particles and
  thus the prevailing mode structure \cite{GK92,Jiang}. (As an
  example, consider a plasma with grains of $R = 1$ micron, mass
  density $\rho \sim 1.5$ g/cm$^3$ and $Z \sim 3000$; here a magnetic
  field of 2 Teslas would produce a dust cyclotron frequency
  $\omega_{\rm cd} \approx 0.16$ rad/s, and with $T_{\rm d} \sim 0.03$
  eV a gyroradius $r_{\rm gyro} \approx 0.5$ cm. Thus, based on the
  criterion $r_{\rm gyro} <$ confinement length, the creation of a
  magnetized plasma may become feasible. A more restrictive criterion
  may, however, emerge from the requirement $\omega_{\rm cd} >
  \nu_{\rm coll}$, the grain-neutral collision frequency. In a
  different vein, systems containing magnetically polarizable plasma
  particles are expected to exhibit a series of novel physical
  phenomena, both in equilibrium \cite{Feldman07} and in terms of
  collective excitations \cite{Feldman06}.}
\item{ 
  Transport coefficients may have been a legitimate topic for
  consideration in this Review, but partly for reason of economy,
  partly because their treatment requires a different (theoretical and
  simulation) methodology from those appropriate for the study of wave
  phenomena, the subject has not been included. The transport
  coefficients of 3D Yukawa systems in the liquid phase are relative
  well known. The self-diffusion was studied in \cite{OH2000},
  estimates for the viscosity were given in
  \cite{Murillo2000}. Molecular dynamics simulations have proven to be
  invaluable tools for studies of transport phenomena and made
  possible the determination of shear viscosity and thermal
  conductivity \cite{SH02,SM01,SC02,SC03,DH04,DonkoHartmann08}. 
  Recent theoretical work on this topic has focused on the mapping between Yukawa,
  Coulomb and hard-sphere systems \cite{FM03,F04}. The effect of  
  Langevin dynamics on the viscosity of 3D Yukawa systems has been
  studied in \cite{Ramazanov2008}. For recent experimental
  work on 3D systems see e.g. \cite{Vaulina2008}.

  The realization of
  2D complex plasma liquids and the development of modern experimental
  (perturbation and data acquisition) techniques resulted in renewed
  interest of transport properties (which are especially interesting
  due to the controversies about the very existence of transport
  coefficients in low-dimensional systems). During the last few years
  several experimental and simulation studies have been carried out on
  the shear viscosity
  \cite{transport1,transport2,LGV2006,Kazah,PRL2006}, thermal
  conductivity \cite{transport3,heat2008} and diffusion
  \cite{sdiff2007,Goree2008} properties of 2D Yukawa liquids, and this
  topic is expected to attract further attention. }

\end{itemize}

\ack

This work has been partially supported by OTKA-T-48389, MTA-NSF/102,
OTKA-PD-049991, OTKA-IN-69892, NSF PHY-0206695, NSF PHY-0715227,
NSF PHY-0514619 and
DE-FG02-03ER54716 grants. We thank Marlene Rosenberg, Stamatios
Kyrkos, John Goree for numerous helpful contributions without which
this Review would have been less complete, to Kenneth I. Golden for critically 
reading part of the manuscript and to Pradip Bakshi for discussions.

\section*{References}

\bibliography{jpcm}
\bibliographystyle{iopart-num}

\end{document}